\documentclass[preprint,nofootinbib,aps,superscriptaddress,eqsecnum]{revtex4-1} 
\pdfoutput=1
\usepackage{rotating}
 \usepackage{graphicx}
 \usepackage{amsmath}
\usepackage{amsfonts}
\usepackage{subfigure}
\usepackage{amssymb}
\usepackage{hyperref}
\usepackage{gensymb}
\usepackage{tikz-feynman}
\usepackage{slashed}
\usepackage{caption}
\usepackage{ulem}
\def\bea{\begin{eqnarray}}
\def\eea{\end{eqnarray}}
\def\be{\begin{equation}}
\def\ee{\end{equation}}

\begin{document} 
\title{Gravitational radiation from binary systems in massive graviton theories}
\author{Tanmay Kumar Poddar }
\email[Email Address: ]{tanmay@prl.res.in}
\affiliation{Theoretical Physics Division, 
Physical Research Laboratory, Ahmedabad - 380009, India}
\author{Subhendra Mohanty}
\email[Email Address: ]{mohanty@prl.res.in}
\affiliation{Theoretical Physics Division, 
Physical Research Laboratory, Ahmedabad - 380009, India}
\author{Soumya Jana}
\email[Email Address: ]{soumyajana.physics@gmail.com}
\affiliation{Department  of  Physics,  Sitananda  College,  Nandigram,  721631,  India}
\begin{abstract}
Theories with massive gravitons have peculiarity called the van Dam-Veltman-Zakharov discontinuity in that the massive theory propagator does not go to the massless graviton propagator in the zero graviton mass limit. This results in large deviation in Newtons law for massive graviton theories even when the graviton mass vanishes. We test the vDVZ in massive graviton theories for single graviton vertex process namely the gravitational radiation from a classical source.     
We calculate the gravitational radiation from compact binaries using  the perturbative Feynman diagram method. We perform this calculation for Einstein's gravity with massless gravitons and verify that the Feynman diagram calculation reproduces the quadrupole formula. Using the same procedure we calculate the gravitational radiation for three massive graviton theories: (1) the Fierz-Pauli theory (2) the modified Fierz-Pauli theory without the vDVZ discontinuity and (3) the Dvali-Gabadadze-Porrati theory with a momentum dependent graviton mass. We put limits on the graviton mass in each of these theories from observations of binary pulsar timings.

\end{abstract}

\maketitle
\section{Introduction}
 
 Einstein's general relativity (GR), since its inception in 1916, has passed all experimental tests \cite{Weinberg:1972kfs}. To move towards the correct  quantum  theory of gravity, it is important to test  which variations of classical GR  fail the experimental tests or have some theoretical inconsistencies. One such variation of GR which has been widely studied is the Fierz-Pauli (FP) theory of massive gravity \cite{Fierz:1939ix}. In a scalar or vector field theory a massive particle exchange gives rise to a $(1/r)e^{-m_g r}$ Yukawa potential which goes to the $1/r$ potential in the $m_g\rightarrow 0$ limit.    
   The FP theory of massive graviton has the peculiarity that in the zero graviton mass limit the Lagrangian goes smoothly to Einstein-Hilbert (EH) linearized gravity theory, while the graviton propagator has additional contributions from the scalar modes of the metric which do not decouple in the zero graviton mass limit. As a result, the Newtonian potential in the zero-mass limit of FP theory is a factor $(4/3)$ larger than the prediction from the EH theory (which of course agrees with the Newtonian potential).
 This peculiarity of the FP theory where the action goes to the EH theory in the zero mass limit but the graviton propagator does not was first pointed out by van Dam and Veltman \cite{vanDam:1970vg} and independently by Zakharov \cite{Zakharov:1970cc} and this feature which arises in most massive gravity theories \cite{Hinterbichler:2011tt,deRham:2014zqa, Mitsou:2015yfa,Joyce:2014kja} is called the van Dam-Veltman-Zakharov (vDVZ) discontinuity (however, in the nonlinear FP theory, a proper decoupling limit will display the vDVZ discontinuity already in the action). Experimental constraints on the graviton mass are listed in \cite{deRham:2016nuf}.
 
 It is of interest to ask if instead of a graviton exchange diagram  we consider a  one graviton vertex process like gravitational radiation then, whether there is a difference in the result between the predictions of GR and the predictions of massive gravity theories in the $m_g\rightarrow 0$ limit and whether a manifestation of the vDVZ discontinuity can be seen in this phenomenon.
  
 GR in the weak field limit can be treated as a quantum field theory of spin-2 fields in the Minkowski space \cite{Feynman:1996kb, Weinberg:1964ew,Veltman:1975vx, Donoghue:2017pgk,Kuntz:2019zef}.  Any classical gravity interaction like Newtonian potential between massive bodies or bending of light by a massive body can be described by a tree level graviton exchange diagram. The result of the tree level diagrams should match the weak field classical GR results.
 The derivation of gravitational radiation from binary stars as a single vertex Feynman diagram of massless graviton emission from a classical source has been performed in \cite{Mohanty:1994yi,Mohanty:2020pfa} and the results match with the result of Peter and Mathews  \cite{Peters:1963ux} who used the quadrupole formula of classical GR. 
 
 The first evidence of Gravitational Wave (GW) radiation was obtained from precision measurements of the Hulse -Taylor binary system \cite{Hulse:1974eb,Taylor:1982zz,Weisberg:1984zz}. The orbital period loss of the compact binary system confirms Einstein's GR  \cite{Peters:1963ux} to $\sim 0.1 \% $ accuracy \cite{Weisberg:2016jye}. Following the Hulse-Taylor binary there have been other precision measurements from compact binary systems \cite{Kramer:2006nb,Antoniadis:2013pzd,Freire:2012mg}. 

 Binary stars can also radiate other ultra-light fields like axions or gauge bosons. The angular frequency of pulsar binaries is $\Omega \sim 10^{-19}$ eV and particles with a mass lower than $\Omega$ can be radiated like the radiation of gravitational waves. The Feynman diagram method is pedagogically simpler to generalize the calculation of scalars and gauge bosons. A calculation of radiation of ultra-light scalars \cite{Mohanty:1994yi}, axions \cite{Poddar:2019zoe}, and gauge bosons \cite{Poddar:2019wvu} has been performed with this method and compared with experimental observations of binary pulsars (or pulsar-white dwarf binaries).  This enables us to probe the couplings of ultra-light dark matter \cite{Hu:2000ke, Hui:2016ltb} which are predicted to be in the mass range  $\sim  10^{-21}-10^{-22}$ eV  to be probed with binary pulsar timing measurements.
  
  In this paper, we study massive graviton theories with a single vertex process namely graviton radiation from binary stars and  we consider three models (1) the Fierz-Pauli ghost free theory which has a vDVZ discontinuity in the propagator, (2) a modification of Fierz-Pauli theory where there is a cancellation between the ghost and the scalar degrees so that there is no vDVZ discontinuity \cite{Visser:1997hd,Finn:2001qi,Gambuti:2020onb,Gambuti:2021meo} and (3) the Dvali-Gabadadze-Porrati (DGP) theory \cite{ Dvali:2000hr, Dvali:2000rv,Dvali:2000xg} which is  ghost-free but the extra scalar degree of freedom gives rise to the vDVZ discontinuity. The mass term in DGP gravity is momentum dependent which serves the purpose of suppressing the long range interactions in a virtual graviton exchange process.  For real gravitons the graviton mass is tachyonic. We compare our results with observations and put limits on the graviton mass allowed in each of these theories.

    We also compare our results  with the earlier classical field calculations in massive gravity theories \cite{VanNieuwenhuizen:1973qf, Will:1997bb, Larson:1999kg,Finn:2001qi, deRham:2012fw, Shao:2020fka}. There are several existing bounds on graviton mass considering the tests of Yukawa potential, from modified dispersion relation, fifth force constraints, etc. (see \cite{deRham:2016nuf} for review). In particular, considering Vainshtein screening at the non-linear scales of the massive theories of gravity, measurements have already ruled out a range of $m_g$ below the Vainshtein threshold in various systems. For example, from the Lunar Laser ranging experiments for the Earth-Moon system, the graviton mass range $10^{-32}$ eV $<m_g< 10^{-20}$ eV is ruled out \cite{Dvali:2002vf}. For any theory
containing the cubic Galileon in the decoupling limit (i.e. the Vainshtein screened regime), from the Hulse-Taylor pulsar the mass range $10^{-27}$ eV $< m_g < 10^{-24}$ eV is ruled out \cite{deRham:2012fw}. In this paper, we investigate the complementary regime, i.e. the unscreened linear regime and, hence the mass ranges greater than the Vainshtein threshold value for the binary pulsar systems.
    
   This paper is organized as follows. In Section \ref{FP} we discuss the Fierz-Pauli theory and derive the  formula for energy loss by graviton radiation using the Feynman diagram  method. In Section \ref{FPa} we do the same study for the modified FP theory without the vDVZ discontinuity and in Section \ref{DGP} we study the DGP theory. In Section \ref{Observations}, we compare the results with observations from the Hulse-Taylor binary (PSR B1913+16) and pulsar white dwarf binary (PSR J1738+0333) and put limits on the graviton mass for each of the massive gravity theories discussed. We also discuss the limits of applicability of the perturbation theory from the Vainshtein criterion  and the corresponding limits on the range of graviton mass established from binary stars. In Section \ref{Conclusions}, we summarise the results and discuss future directions. In Appendix \ref{appendi}, we give the detailed derivation of the Feynman diagram method of calculating gravitational radiation from compact binaries in GR for comparison with massive gravity theory results discussed in this paper.

\section{Fierz-Pauli massive gravity theory}
\label{FP}

The Fierz-Pauli theory \cite{Fierz:1939ix} is {described by the action 
\bea
S&=&\int d^4x\Big[ -\frac{1}{2} (\partial_\mu h_{\nu \rho})^2 + \frac{1}{2} (\partial_\mu h)^2 
- (\partial_\mu h)(\partial^\nu h^\mu_\nu)+(\partial_\mu h_{\nu \rho} )(\partial^\nu h^{\mu \rho})\nonumber\\
&&\quad \quad \quad\quad+\frac{1}{2}m^2_g\Big(h_{\mu\nu}h^{\mu\nu}-h^2\Big)+\frac{\kappa}{2}h_{\mu\nu}T^{\mu\nu} \Big] \nonumber\\
&=&\int d^4x \left[ \frac{1}{2} h_{\mu \nu} {\cal E}^{\mu \nu \alpha \beta}h_{\alpha \beta} 
+\frac{1}{2}m^2_g h_{\mu \nu} (\eta^{\mu (\alpha } \eta^{\beta) \nu}- \eta^{\mu \nu}\eta^{\alpha \beta})h_{\alpha \beta}+ \frac{\kappa}{2} h_{\mu \nu }T^{\mu \nu} \right],
\label{eq:FP_action}
\eea
where the operator $ {\cal E}^{\mu \nu \alpha \beta}$ is given in  Eq.\ref{kineticOp}. The mass term breaks the gauge symmetry $h_{\mu \nu} \rightarrow h_{\mu \nu} - \partial_\mu \xi_\nu -\partial_\nu \xi_\mu$. We will assume that the energy-momentum is conserved, $\partial_\mu T^{\mu \nu}=0$.

The equation of motion from Eq.\ref{eq:FP_action} is
\bea
\left( \Box +m_g^2 \right) h_{\mu \nu} -\eta_{\mu \nu} \left(\Box +m_g^2\right) h - \partial_\mu \partial^\alpha h_{\alpha \nu} - \partial_\nu \partial^\alpha h_{\alpha \mu} + \eta_{\mu \nu} \partial^\alpha \partial^\beta h_{\alpha \beta}+ \partial_\mu \partial_\nu h = -\kappa T_{\mu \nu}. \nonumber\\
\label{eom1}
\eea
Taking the divergence of Eq.\ref{eom1} we have
\be
m_g^2 \left( \partial^\mu h_{\mu \nu} -\partial_\nu h \right)=0.
\label{eom2}
\ee
These are 4 constraint equations which reduce the independent degrees of freedom of the graviton from 10 to 6. 

Using Eq.\ref{eom2} in Eq.\ref{eom1} we obtain 
\be
\Box h_{\mu \nu} -\partial_\mu \partial_\nu h +m_g^2\left(h_{\mu \nu} -\eta_{\mu \nu}h\right)= -\kappa T_{\mu \nu}.
\ee
Taking the trace of this equation we obtain the relation
\be
h=\frac{\kappa }{3 m_g^2} T.
\ee
Therefore trace $h$ is not a propagating mode but is determined algebraically from the trace of the stress tensor. This is the ghost mode as the kinetic term for $h$ in Eq.\ref{eom1} appears with the wrong sign. Therefore in the Fierz-Pauli theory the ghost mode does not propagate. The number of independent propagating degrees of freedom of the Fierz Pauli theory is therefore 5. These are 2 tensor modes, 2 three-vector degrees of freedom which do not couple to the energy-momentum tensor  and 1 scalar which couples to the trace of the energy-momentum tensor.

The propagator in the FP theory is given formally by 
\be
\left[ {\cal E}^{\mu \nu \alpha \beta}+m_g^2 \left(\eta^{\mu (\alpha } \eta^{\beta) \nu}- \eta^{\mu \nu}\eta^{\alpha \beta} \right)\right] D^{(m)}_{\alpha \beta \rho \sigma}(x-y)=\delta^\mu_{(\rho}\delta^\nu_{\sigma)}\delta^4(x-y).
\label{Dm1}
\ee
Going to momentum space ($\partial_\mu \rightarrow i k_\mu$) we can find $D^{(m)}_{\alpha \beta \rho \sigma}(k)$ from Eq.\ref{Dm1}. The propagator for the Pauli-Fierz massive graviton turns out to be
\be
D^{(m)}_{\alpha \beta \rho \sigma}(k)=  \frac{1}{-k^2 +m_g^2} \left (\frac{1}{2} (P_{\alpha \rho}  P_{\beta \sigma} + P_{\alpha \sigma }P_{\beta \rho}) - \frac{1}{3} P_{\alpha  \beta}P_{\rho \sigma}\right),
\label{Dm2}
\ee
where 
\be
P_{\alpha \beta}\equiv\eta_{\alpha \beta}- \frac{k_\alpha k_\beta}{m_g^2}.
\ee

In tree level processes where there is a graviton exchange between conserved currents, the  amplitude is of the form 
\be
{\cal A}_{FP} =\frac{\kappa^2}{4}  T^{\alpha \beta} D^{(m)}_{\alpha \beta \mu \nu}T^{\prime \mu \nu}.
\label{ampFP}
\ee
The momentum dependent terms will vanish due to conservation of the stress tensor $k_\mu T^{\mu \nu}= k_\nu T^{\mu \nu}=0$. Hence, for tree level calculations one may drop the momentum dependent terms in Eq.\ref{Dm2} and the propagator for the FP theory may be written as
\be
D^{(m)}_{\mu \nu \alpha \beta }(p)=  \frac{1}{-k^2 +m_g^2} \left (\frac{1}{2} (\eta_{\alpha \mu}  \eta_{\beta \nu} + \eta_{\alpha \nu }\eta_{\beta \mu}) - \frac{1}{3} \eta_{\alpha  \beta}\eta_{\mu \nu}  + (k{\rm -dependent\, terms}) \right).
\label{Dm3}
\ee
When the graviton is treated as a quantum field, the Feynman propagator is defined as in the massless theory Eq.\ref{prop2},
\bea
D^{(m)}_{\mu \nu \alpha \beta}(x-y)&=& \langle 0 | T (\hat {h}_{\mu \nu}(x) \hat {h}_{\alpha \beta}(y))|0\rangle\nonumber\\
&=&\int \frac{d^4 k}{(2 \pi)^4}\frac{1}{-k^2 + m_g^2 + i \epsilon} e^{i k(x-y)} 
\sum_{\lambda} \epsilon_{\mu\nu}^\lambda(k)\epsilon_{ \alpha\beta}^{*\lambda}(k).
\label{Dm4}
\eea
Comparing Eq.\ref{Dm3} and Eq.\ref{Dm4} we see that the polarisation sum for the FP massive gravity theory can be written as
\be
\sum_{\lambda} \epsilon_{\mu\nu}^\lambda(k)\epsilon_{ \alpha\beta}^{*\lambda}(k)= \frac{1}{2} (\eta_{\mu \alpha}  \eta_{\nu \beta} + \eta_{\nu \alpha }\eta_{\mu \beta }) - \frac{1}{3} \eta_{\alpha  \beta}\eta_{\mu \nu}  + (k{\rm -dependent\, terms}).
\label{polsumFP}
\ee

In processes where there is graviton emission from an external leg as in the case of gravitational wave radiation from a classical current, the amplitude square will have the form
\be
|{\cal M}|^2 = \left(\frac{\kappa^2}{4} \right )\sum_\lambda  | \epsilon_{\mu\nu}^\lambda(k) T^{\mu\nu}(k^\prime)|^2=  \left(\frac{\kappa^2}{4} \right ) \sum_\lambda \epsilon_{\mu\nu}^\lambda(k)\epsilon_{ \alpha\beta}^{*\lambda}(k) T^{\mu\nu}(k^\prime) T^{*\alpha \beta}(k^\prime).
\ee

Since $T^{\mu \nu}$ is a conserved current, the momentum dependent pieces in the polarisation sum will give zero and we can drop them from Eq.\ref{polsumFP} for the  calculations of diagrams with graviton emission from external legs as we will do in this paper.

We see that when the  propagator Eq.\ref{prop1} and polarisation sum Eq.\ref{app3} of the massless graviton theory is compared with the corresponding quantities Eq.\ref{Dm1} and Eq.\ref{polsumFP}, the massive theory differs from the massless theory even in the $m_g\rightarrow 0$ limit. There is an extra contribution of $(1/6) T^* T^\prime$ to the amplitude Eq.\ref{ampFP} in the FP theory. This is the contribution of the scalar degree of freedom of $g_{\mu \nu}$ which does not decouple in the $m_g\rightarrow 0$ limit. 

Consider the Newtonian potential between two massive bodies. The amplitude for the diagram with one graviton exchange is in GR is
\be 
{\cal A}_{GR}= \frac{\kappa^2}{4} T^{\mu \nu} D^{(0)}_{\mu \nu \alpha \beta}(k)T^{\prime \alpha \beta}\,.
\label{amp1}
\ee
The stress tensor for massive bodies at rest in a given reference frame is of the form $T^{\mu \nu}= (M_1, 0,0,0)$ and $T^{\prime \alpha \beta}= (M_2, 0,0,0)$ and the massless graviton propagator in GR is Eq.\ref{prop1}. The  potential derived from Eq.\ref{amp1} is the usual Newtonian form
\bea
V_{GR}&= &\frac{\kappa^2}{4}\int\frac{d^3k}{(2\pi)^3} e^{i k\cdot r}\frac{1}{-k^2} \left(T_{\mu \nu}-\frac{1}{2} \eta_{\mu \nu} T^{\alpha}_{\alpha}\right) T^{\prime \mu \nu}\nonumber\\
&=& \frac{G M_1 M_2}{r},
\eea
where $\kappa=\sqrt{32\pi G}$, and $G$ stands for universal gravitational constant. On the other hand in the Fierz-Pauli theory  the one graviton exchange amplitude Eq.\ref{ampFP} is 
\be
{\cal A}_{FP} = \frac{\kappa^2}{4}\frac{1}{-k^2+m_g^2} \left(T_{\mu \nu}-\frac{1}{3} \eta_{\mu \nu} T^{\alpha}_{\alpha}\right) T^{\prime \mu \nu},
\ee
and the gravitational potential between two massive bodies in the FP theory is 
\bea
V_{FP}= &= &\frac{\kappa^2}{4}\int\frac{d^3k}{(2\pi)^3} e^{i k\cdot r}\frac{1}{-k^2+m_g^2} \left(T_{\mu \nu}-\frac{1}{3} \eta_{\mu \nu} T^{\alpha}_{\alpha}\right) T^{\prime \mu \nu}\nonumber\\
&=& \left( \frac{4}{3}\right) \frac{G M_1 M_2}{r}e^{-m_g r}.
\label{VFP}
\eea

The FP theory gives a Yukawa potential as expected however, in the $m_g \rightarrow 0$ limit the gravitational potential between massive bodies in the FP theory is a factor $4/3$ larger than the Newtonian potential arising from GR.  This is ruled out from solar system tests of gravity \cite{Talmadge:1988qz} even in the $m_g\rightarrow 0$ limit. We note here that the bending of light by massive bodies is unaffected (in $m_g\rightarrow 0$  limit) as the stress tensor for photons $T^\mu_\nu=(\omega,0,0,-\omega)$ is traceless and the scattering amplitudes ${\cal A}_{FP}(m_g\rightarrow 0)= {\cal A}_{GR}$. Experimental observations \cite{Fomalont:2009zg} of the bending of radio waves by the Sun matches GR to 1\%. The two observations together imply that the extra factor of (4/3) in the Newtonian potential of FP theory cannot be absorbed by redefining $G$.

The fact that the action of the FP theory \ref{eq:FP_action} goes to the Einstein-Hilbert action Eq.\ref{EH} in the $m_g \rightarrow 0$ limit while the propagator Eq.\ref{Dm3} does not go to the massless form Eq.\ref{prop1}, is what is called the vDVZ discontinuity  pointed out by van Dam and Veltman \cite{vanDam:1970vg} and Zakharov \cite{Zakharov:1970cc}. 

It has been pointed out by Vainshtein \cite{Vainshtein:1972sx, Babichev:2013usa} that the linear FP theory breaks down at distances much larger than the Schwarzschild  radius $R_s=2GM$ below which the linearised GR is no longer valid ($\kappa h_{\mu \nu} \sim 1$ (below this distance)). The scalar mode in FP theory becomes strongly coupled with decreasing $m_g$ and the minimum radius from a massive body at which the linearised FP theory is valid is called the Vainshtein radius and is given by $R_V = (R_s /m_g^4)^{1/5}$. We will discuss the Vainshtein radius of different theories of gravity discussed in this paper and how this consideration limits the bounds on $m_g$ from binary systems derived in this paper in Section\ref{Vainshtein}.

\subsection{Graviton radiation from binaries  in Fierz-Pauli theory}
\begin{figure}
\centering
\includegraphics[width=4.0in,angle=360]{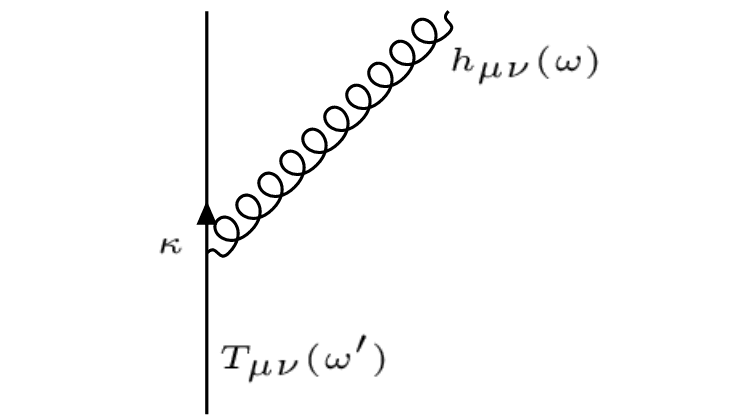}
\caption{Emission of graviton from a classical source.}
\label{fig:feyn}
\end{figure}
We consider the graviton radiation from the compact binary systems classically. The pictorial representation of graviton emission from a classical source is shown in FIG.\ref{fig:feyn}. The classical graviton current $T^{\mu\nu}$ is detemined from the Kepler's orbit and the interaction vertex is $\frac{1}{2}\kappa h_{\mu\nu}T^{\mu\nu}$, where $h_{\mu\nu}$ is the graviton field and $\kappa=\sqrt{32\pi G}$. Here we use linearized gravity formulation with an extension of non zero graviton mass term  Eq.\ref{eq:FP_action} to calculate the energy loss of a compact binary system due to graviton emission.

The emission rate of graviton from the interaction Lagrangian between the gravity and source $(\frac{1}{2}\kappa h_{\mu\nu}T^{\mu\nu})$ is given by
\begin{equation}
d\Gamma= \frac{\kappa^2}{4}\sum_{\lambda} |T_{\mu\nu}(k^\prime)\epsilon^{\mu\nu}_\lambda(k)|^2 2\pi \delta(\omega-\omega^\prime)\frac{d^3k}{(2\pi)^3}\frac{1}{2\omega},
\label{eq:1}
\end{equation}
where $T_{\mu\nu}(k^\prime)$ is the classical graviton current in the momentum space. Expanding the modulus squared in Eq.\ref{eq:1}, we can write
\begin{equation}
d\Gamma=\frac{\kappa^2}{8(2\pi)^2}\sum_{\lambda}\Big(T_{\mu\nu}(k^\prime)T^*_{\alpha\beta}(k^\prime)\epsilon^{\mu\nu}_\lambda(k)\epsilon^{*\alpha\beta}(k)\Big)\frac{d^3k}{\omega}\delta(\omega-\omega^\prime).
\label{eq:2}
\end{equation}
Using the  polarization sum of the Fierz-Pauli theory  \ref{polsumFP} this expression becomes
\begin{eqnarray}
d\Gamma&=&\frac{\kappa^2}{8(2\pi)^2}\int \Big[T_{\mu\nu}(k^\prime)T^*_{\alpha\beta}(k^\prime)\Big]\Big[\frac{1}{2}(\eta^{\mu\alpha}\eta^{\nu\beta}+\eta^{\mu\beta}\eta^{\nu\alpha}-\eta^{\mu\nu}\eta^{\alpha\beta})+\frac{1}{6}\eta^{\mu\nu}\eta^{\alpha\beta}\Big]\frac{d^3k}{\omega}\delta(\omega-\omega^\prime)\,.\nonumber\\
\eea
The extra $(1/6)\eta^{\mu\nu}\eta^{\alpha\beta}$  term compared to the massless graviton case is the contribution of the scalar mode in FP theory. Simplifying, we obtain
\bea
d\Gamma&=&\frac{\kappa^2}{8(2\pi)^2}\int \Big[|T_{\mu\nu}(k^\prime)|^2-\frac{1}{3}|T^{\mu}{}_{\mu}(k^\prime)|^2\Big]\delta(\omega-\omega^\prime)\omega \Big(1-\frac{m^2_{g}}{\omega^2}\Big)^\frac{1}{2}d\omega d\Omega_k,
\label{eq:4}
\end{eqnarray}
where we have used $d^3k=k^2dkd\Omega$ and the dispersion relation $k^2=(\omega^2-m^2_g)$. From the emission rate we can calculate the rate of energy loss due to massive graviton emission which is 
\begin{equation}
\frac{dE}{dt}=\frac{\kappa^2}{8(2\pi)^2}\int \Big[|T_{\mu\nu}(k^\prime)|^2-\frac{1}{3}|T^{\mu}{}_{\mu}(k^\prime)|^2\Big]\delta(\omega-\omega^\prime)\omega^2 \Big(1-\frac{m^2_{g}}{\omega^2}\Big)^\frac{1}{2}d\omega d\Omega_k.
\label{eq:5}
\end{equation}

For the massive graviton, the dispersion relation is 
\begin{equation}
|\textbf{k}|^2=\omega^2\Big(1-\frac{m^2_g}{\omega^2}\Big).
\label{eq:a1}
\end{equation}
Hence, the unit vector along the momentum direction of graviton is $\hat{k^i}=\frac{k^i}{\omega\sqrt{1-\frac{m^2_g}{\omega^2}}}$. Using the relation $k_\mu T^{\mu\nu}=0$ and Eq.\ref{eq:a1}, we can write the $T_{00}$ and $T_{i0}$ components of the stress tensor in terms of $T_{ij}$ as follows
\begin{equation}
T_{0j}=-\sqrt{1-\frac{m^2_g}{\omega^2}}\hat{k^i}T_{ij},\hspace{0.5cm} T_{00}=\Big(1-\frac{m^2_g}{\omega^2}\Big)\hat{k^i}\hat{k^j}T_{ij}.
\end{equation}
Hence, we can write
\begin{equation}
\Big[|T_{\mu\nu}(k^\prime)|^2-\frac{1}{3}|T^{\mu}{}_{\mu}(k^\prime)|^2\Big]\equiv {\Lambda_{ij,lm}}T^{ij*}T^{lm},
\end{equation}
where,
\begin{equation}
{\Lambda_{ij,lm}}=\Big[\delta_{il}\delta_{jm}-2\Big(1-\frac{m^2_g}{\omega^2}\Big)\hat{k_j}\hat{k_m}\delta_{il}+\frac{2}{3}\Big(1-\frac{m^2_g}{\omega^2}\Big)^2\hat{k_i}\hat{k_j}\hat{k_l}\hat{k_m}-\frac{1}{3}\delta_{ij}\delta_{lm}+\frac{1}{3}\Big(1-\frac{m^2_g}{\omega^2}\Big)\Big(\delta_{ij}\hat{k_l}\hat{k_m}+\delta_{lm}\hat{k_i}\hat{k_j}\Big)\Big].
\label{rat1}
\end{equation}
Therefore, we can write Eq.\ref{eq:5} as
\begin{equation}
\frac{dE}{dt}=\frac{\kappa^2}{8(2\pi)^2}\int {\Lambda_{ij,lm}}T^{ij*}T^{lm} \delta(\omega-\omega^\prime)\omega^2 \Big(1-\frac{m^2_{g}}{\omega^2}\Big)^\frac{1}{2}d\omega d\Omega_k.
\end{equation}
We can do the angular integrals using the relations \ref{dOmegak} and obtain,
\begin{eqnarray}
\int d\Omega_k \Lambda_{ij,lm}T^{ij*}(\omega^\prime)T^{lm}({\omega^\prime})&=& \frac{8\pi}{5}\left(\left[\frac{5}{2}-\frac{5}{3}\left(1-\frac{m_g^2}{\omega'^2}\right)+\frac{2}{9}\left(1-\frac{m_g^2}{\omega'^2}\right)^2\right]T^{ij}T*_{ij}\right.\nonumber\\
&& \left. +\left[-\frac{5}{6}+\frac{5}{9}\left(1-\frac{m_g^2}{\omega'^2}\right)+\frac{1}{9}\left(1-\frac{m_g^2}{\omega'^2}\right)^2\right]\vert T^i{}_{i}\vert^2\right).
\label{eq:a2}
\end{eqnarray}

Hence, the rate of energy loss becomes

\begin{eqnarray}
\frac{dE}{dt}&=& \frac{8G}{5}\int \left[\left\lbrace\frac{5}{2}-\frac{5}{3}\left(1-\frac{m_g^2}{\omega'^2}\right)+\frac{2}{9}\left(1-\frac{m_g^2}{\omega'^2}\right)^2\right\rbrace T^{ij}T^*_{ij}\right.\nonumber\\
&& \left. +\left\lbrace -\frac{5}{6}+\frac{5}{9}\left(1-\frac{m_g^2}{\omega'^2}\right)+\frac{1}{9}\left(1-\frac{m_g^2}{\omega'^2}\right)^2\right\rbrace\vert T^i{}_{i}\vert^2\right]\delta(\omega-\omega^\prime)\omega^2 \Big(1-\frac{m^2_{g}}{\omega^2}\Big)^\frac{1}{2}d\omega. \nonumber\\
\label{eq:a4}
\end{eqnarray}
In the massless gravity theory the prefactors of $T^{ij}T^*_{ij}$ and $\vert T^i{}_i\vert^2$ are $1$ and $-1/3$ respectively. Note that the $m_g\rightarrow 0$ limit of Eq.\ref{eq:a4}  gives different prefactors. In the massive graviton limit, all the five polarization components contribute to the energy loss instead of two as in the massless limit. Therefore, from Eq.\ref{eq:a4}, we will not obtain the energy loss for massless limit by simply putting $m_g\rightarrow 0$. In Appendix\ref{appendi} we obtain the energy loss due to massless graviton radiation from compact binary systems. In massive gravity theories, the Newtonian gravitational potential takes different form than GR. As a result the Keplerian orbits are also affected. For, FP theory the potential energy for binary system takes the form of Yukawa-type with 4/3 extra pre-factor as discussed in Eq.~(\ref{VFP}) when there is no screening. However, for GW emission we must have $n_0=m_g/\Omega<1$ which implies that $a<R_V$ and therefore the Newtonian potential for orbital motion of the binary system is Vainshtein screened.  There will be the corrections in the Newtonian gravitational potential energy from the screened scalar mode.\\
Concretely, to see the effects of the scalar
polarisation in this $a < R_V$ limit one can split the massive $h$ into $\tilde{h}+\partial A/m_g + \partial \partial \phi/m_g^2$ such that $\tilde{h}_{\mu\nu}$ now enjoys a gauge invariance and carries only the two tensor modes, while $\phi$ carries the scalar mode (the vector mode $A_{\mu}$ can be consistently set to zero for this matter
configuration).  After $\tilde{h}_{\mu\nu}$ and $\phi$, the action in the decoupling limit is \cite{Arkani-Hamed:2002bjr},
\begin{equation}
S=\int d^4x \left[ \frac{1}{2}\tilde{h}_{\mu\nu}\mathcal{E}^{\mu\nu\alpha\beta}\tilde{h}_{\alpha\beta} -\frac{1}{2}\phi\Box\phi + \frac{1}{2M_{pl}} \tilde{h}_{\mu\nu}T^{\mu\nu}+ \frac{1}{2M_{pl}}\phi T+ \mathcal{L}_{\rm{int}}\right]
\label{action_decoupling}
\end{equation}
The precise interactions will depend on specific massive gravity theory. For FP theory,  there will be non-linearities like,
\begin{equation}
\mathcal{L}_{\rm{int}}\sim \left[\alpha(\Box \phi)^3+\beta \Box\phi \phi_{,\mu\nu}\phi^{,\mu\nu}\right],
\end{equation} 
where $\alpha$ and $\beta$ are model dependent coefficients. At $r=a<<R_V$, deep inside the Vainshtein region, the equation of motion for $\phi$ gives,
\begin{eqnarray*}
\frac{\phi}{M_{pl}}\sim m_g^2\sqrt{R_sa^3}\sim n_0\frac{h}{M_{pl}},
\end{eqnarray*}
from balancing $\mathcal{L}_{\rm{int}}\sim \phi^3/(M_{pl}m_g^4r^6)$ against $\phi T/M_{pl} \sim \phi M/(M_{pl}r^3)$. Here $a$ denotes the semi major axis of the binary orbit. So the scalar fifth force is suppressed by $n_0$ relative to the Newtonian force. \\
However, we neglect the corrections as they are small and will not affect our order of magnitude results and, therefore, we only consider the GW stress-energy tensor. Thus our results are approximate and not valid for all orders of $n_0$.


From Eq.\ref{eq:13} we get
\begin{equation}
\Big[T_{ij}(\omega^\prime)T^*_{ji}(\omega^\prime)-\frac{1}{3}|T^{i}{}_{i}(\omega^\prime)|^2\Big]=4\mu^2{\omega^{\prime}}^4a^4f(n,e).
\label{s1}
\end{equation}
where $n_0=\frac{m_g}{\Omega}$, and 
\begin{equation}
\begin{split}
f(n,e)=\frac{1}{32n^2}\Big\{[J_{n-2}(ne)-2eJ_{n-1}(ne)+2eJ_{n+1}(ne)+\frac{2}{n}J_n(ne)-J_{n+2}(ne)]^2+\\
(1-e^2)[J_{n-2}(ne)-2J_n(ne)+J_{n+2}(ne)]^2+\frac{4}{3n^2}J^2_{n}(ne)\Big\}.
\end{split}
\end{equation}

The final expression of $dE/dt$ for massive Fierz Pauli theory can be written in the compact form as
\begin{equation}
\begin{split}
\frac{dE}{dt}= \frac{32G}{5} \mu^2 a^4\Omega^6 \sum_{n=1}^{\infty} n^6\sqrt{1-\frac{n_0^2}{n^2}}\left[f(n,e)\left( \frac{19}{18}+\frac{11}{9}\frac{n_0^2}{n^2}+\frac{2}{9}\frac{n_0^4}{n^4}\right) + \frac{5J^2_n(ne)}{108n^4}\left(1-\frac{n_0^2}{n^2}\right)^2\right].
\end{split}
\label{eq:dedt_FP}
\end{equation}
We can split Eq.\ref{eq:dedt_FP} as
\begin{equation}
\begin{split}
\frac{dE}{dt}= \frac{32G}{5} \mu^2 a^4\Omega^6 \sum_{n=1}^{\infty} n^6\sqrt{1-\frac{n_0^2}{n^2}}\left[f(n,e)\left( 1+\frac{4}{3}\frac{n_0^2}{n^2}+\frac{1}{6}\frac{n_0^4}{n^4}\right) -\frac{5J^2_n(ne)}{36n^4}\frac{n_0^2}{n^2}\left(1-\frac{n_0^2}{4n^2}\right)\right]+\\
\frac{32G}{5} \mu^2 a^4\Omega^6 \sum_{n=1}^{\infty} n^6\sqrt{1-\frac{n_0^2}{n^2}}\Big[\frac{1}{18}f(n,e)\Big(1-\frac{n^2_0}{n^2}\Big)^2+\frac{5J^2_n(ne)}{108n^4}\Big(1+\frac{n^2_0}{2n^2}\Big)^2\Big],
\end{split}
\label{eq:wa}
\end{equation}
where the first term in Eq.\ref{eq:wa} denotes the energy loss in the massive gravity theory without vDVZ discontinuity (Eq.\ref{eq:dedt_FPd}) and the second term denotes the contribution due to the scalar mode associated with $\frac{1}{6}\eta_{\mu\nu}\eta_{\alpha\beta}$.
We can also write Eq.\ref{eq:wa} to the leading order in $n^2_0$ as
\begin{equation}
\frac{dE}{dt}\simeq \frac{32G}{5} \mu^2 a^4\Omega^6 \Big[\sum_{n=1}^{\infty}\Big(\frac{19}{18}n^6f(n,e)+\frac{5}{108}n^2J^2_n(ne)\Big)+n^2_0\sum_{n=1}^{\infty}\Big(\frac{25}{36}n^4f(n,e)-\frac{25}{216}J^2_n(ne)\Big)\Big]+\mathcal{O}(n^4_0).
\end{equation}
The rate of energy loss in the Keplerian orbit leads to the decrease in orbital period decay at a rate
\begin{equation}
\dot{P_b}=-6\pi G^{-\frac{3}{2}}(m_1m_2)^{-1}(m_1+m_2)^{-\frac{1}{2}}a^{\frac{5}{2}}\Big(\frac{dE}{dt}\Big).
\label{ms}
\end{equation}

\begin{figure}
\centering
\includegraphics[width=12.0cm]{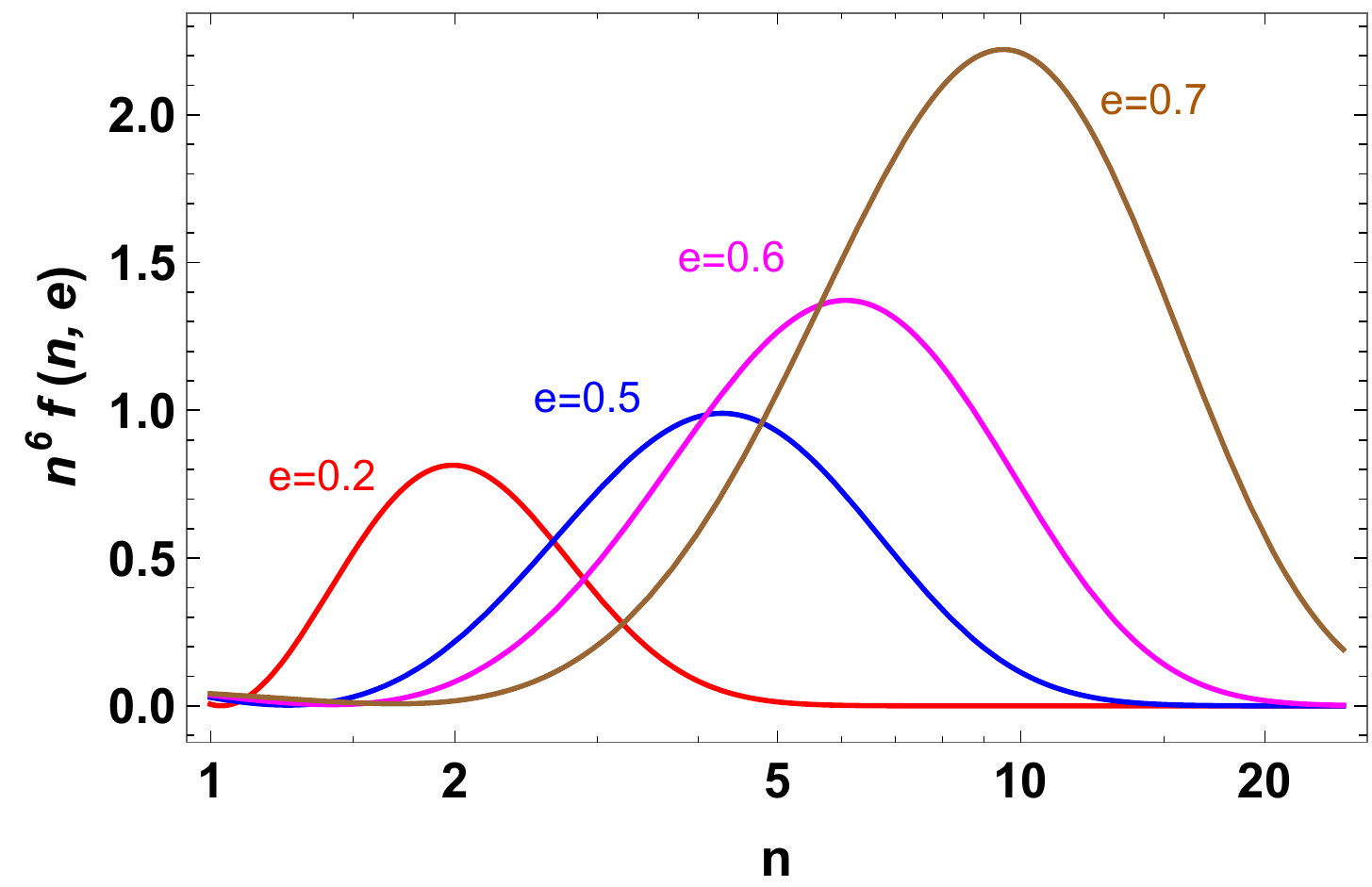}
\caption{Variation of $n^6f(n,e)$ with $n$ for different orbital eccentricity.}
\label{fig:feynu}
\end{figure}
The energy loss or the power radiated from the binary system increases with increasing the eccentricity as it is clear from FIG.\ref{fig:feynu}, since the energy loss in the first term is proprtional to $n^6f(n,e)$. The radiation is dominated by the higher harmonics for $e\approx 1$. The radiation has a peak at some particular value of $n$ for a given eccentric orbit.

\section{Massive gravity without vDVZ discontinuity}\label{FPa}

In the Fierz-Pauli theory Eq.\ref{eq:FP_action}  there is no ghost owing to the fact that  the relative coefficients of the $h^2$ and $h_{\mu \nu}h^{\mu \nu}$ terms is choosen as $-1$. Generalising the theory beyond this point will lead to the appearance of ghosts. There is a special choice of coefficient where the ghost term cancels the extra scalar contribution to the propagator. In this theory therefore there is no vDVZ discontinuity and there are no ghosts \cite{Gambuti:2020onb,Gambuti:2021meo}. Phenomenologically this theory has the simple generalisation of the spin-2 graviton with 2 polarizations which obey  the dispersion relation $k_0^2=|\vec k|^2 +m_g^2$.
Consider the one parameter generalisation of the Fierz-Pauli theory 
\bea
S=\int d^4x \left[ \frac{1}{2} h_{\mu \nu} {\cal E}^{\mu \nu \alpha \beta}h_{\alpha \beta} +\frac{1}{2}m^2_g h_{\mu \nu} \left(\eta^{\mu (\alpha } \eta^{\beta) \nu}-(1-a) \eta^{\mu \nu}\eta^{\alpha \beta})\right) h_{\alpha \beta}+ \frac{\kappa}{2} h_{\mu \nu }T^{\mu \nu} \right],
\label{eq:FP_deformed}
\eea
where $a=0$ corresponds to  the Fierz-Pauli theory Eq.\ref{eq:FP_action}.  We will derive the equations assuming $a\neq 0$ and see which values of $a$ can solve the problem of vDVZ discontinuity which is generic in massive gravity theories.

The equation of motion from Eq.\ref{eq:FP_deformed} is
\bea
\left( \Box +m_g^2 \right) h_{\mu \nu} -\eta_{\mu \nu} \left(\Box +m_g^2(1-a)\right) h - \partial_\mu \partial^\alpha h_{\alpha \nu} - \partial_\nu \partial^\alpha h_{\alpha \mu} + \eta_{\mu \nu} \partial^\alpha \partial^\beta h_{\alpha \beta}+ \partial_\mu \partial_\nu h = -\kappa  T_{\mu \nu}. \nonumber\\
\label{deom1}
\eea
Taking the divergence of Eq.\ref{deom1} we have
\be
m_g^2 \left( \partial^\mu h_{\mu \nu} -(1-a) \partial_\nu h \right)=0.
\label{deom2}
\ee
These are 4 constraint equations which reduce the independent degrees of freedom of the graviton from 10 to 6. 

Using Eq.\ref{deom2} in Eq.\ref{deom1} we obtain 
\be
(\Box +m_g^2) h_{\mu \nu}- a \eta_{\mu \nu} \Box h -(1-2 a)\partial_\mu\partial_\nu h - m_g^2 \eta_{\mu \nu}(1-a)h  = -\kappa T_{\mu \nu}.
\ee
Taking the trace of this equation we obtain 
\be
-2 a \Box h-(3m^2_g -4 m^2_g a)h=- \kappa T.
\ee
We see that the $h$ is now a propagating field if $a \neq 0$. The kinetic term for $h$ appears with a minus sign so $h$ is a ghost field. The homogenous equation for $h$ can be written as
\be
\Box h-m_h^2 h =0
\ee
with the ghost mass given by
\be
m_h^2 = \frac{m^2_g}{2} \left( 1+3 \left(1- \frac{1}{a}\right) \right).
\label{mh}
\ee
The propagator of the deformed Fierz-Pauli  theory  Eq.\ref{eq:FP_deformed} is given \be
\left[ {\cal E}^{\mu \nu \alpha \beta}+m_g^2 \left(\eta^{\mu (\alpha } \eta^{\beta) \nu}- \eta^{\mu \nu}\eta^{\alpha \beta} (1-a) \right)\right] D^{(a)}_{\alpha \beta \rho \sigma}(x-y)=\delta^\mu_{(\rho}\delta^\nu_{\sigma)}\delta^4(x-y).
\label{Dalpha1}
\ee
This equation can be inverted to give the $D^{(a)}_{\alpha \beta \rho \sigma}$ which in momentum space turns out of the form
\bea
D^{(a)}_{\alpha \beta \mu \nu}(k) &=&  \frac{1}{-k^2 +m_g^2} \left (\frac{1}{2} (\eta_{\alpha \mu}  \eta_{\beta \nu} + \eta_{\alpha \nu }\eta_{\beta \mu}) - \frac{1}{3} \eta_{\alpha  \beta}\eta_{\mu \nu} \right) + \frac{i}{k^2 +m_h^2} \left( \frac{1}{6} \eta_{\alpha  \beta}\eta_{\mu \nu} \right ) \nonumber\\ &+& (k{\rm -dependent\, terms}). 
\label{Dalpha2}
\eea
This shows that there are two types of contributions to the propagator helicity-2 states of spin-2 massive gravitons (there are also helicity-1 and helicity-0 states) and a massive scalar with mass $m_h$. This part  is identical to the propagator of the Fierz-Pauli theory. In \ref{Dalpha2} there is an additional contribution from the ghost mode with the kinetic operator $k^2$ with the wrong sign and mass $m_h$ given in Eq.\ref{mh}. The remaining 3 vector degrees of freedom do not couple to the energy momentum tensor and we ignore their contribution here. Now if we choose the parameter $a=1/2$, the mass of the ghost mode Eq.\ref{mh} becomes $m_h^2 = -m_g^2$. The ghost mode for $a=1/2$ becomes tachyonic. Substituting $m_h^2=-m_g^2$ in Eq.\ref{Dalpha2} we see that the propagator simplifies to the form
\bea
D^{(1/2)}_{\alpha \beta \mu \nu}(k) &=&  \frac{1}{-k^2 +m_g^2} \left (\frac{1}{2} (\eta_{\alpha \mu}  \eta_{\beta \nu} + \eta_{\alpha \nu }\eta_{\beta \mu}) - \frac{1}{2} \eta_{\alpha  \beta}\eta_{\mu \nu} \right) + (k{\rm -dependent\, terms}). \nonumber\\
\label{Dalpha3}
\eea
The ghost term with tachyonic mass cancels the extra scalar contribution to the propagator and we are left with the tensor structure of the propagator which is the same as for the massless gravitons Eq.\ref{prop1} but which have the dispersion relations of massive gravitons, $k_0^2= |\vec k|^2 + m_g^2$. In the limit $m_g \rightarrow 0$ the propagator goes to the massles propagator form Eq.\ref{prop1} and thus there is no vDVZ discontinuity. Form the tensor structure of Eq.\ref{Dalpha3} it is clear that for $m_g\rightarrow 0$ the polarisation sum takes the form as Eq.\ref{app3} same as that of the massless theory.

The gravitational potential in this theory takes the Yukawa form
\bea
V^{(1/2)}(r)= \frac{G M_1 M_2}{r}\, e^{-m_g r},
\eea
and the extra factor of (4/3) which was there in the FP theory Eq.\ref{VFP} is absent due to cancellation of the scalar graviton mode with the ghost contribution in the propagator. 
 The Yukawa corrections to the $1/r$ potential will give rise to a perihelion precession in planetary orbits \cite{Poddar:2020exe}. Constraints on the Yukawa potential between planets and the sun which give bounds in the mass of the exchanged particle have been obtained in \cite{Poddar:2020exe}. The long range Yukawa potential caused by axions can also affect the gravitational light bending and Shapiro time delay which is discussed in \cite{KumarPoddar:2021ked}. 

This theory which avoids contributions from the extra scalars mode is phenomenologically the most acceptable. The classical calculation of energy loss from binaries in this spin-2 massive gravity theory was done by Finn and Sutton \cite{Finn:2001qi}. Our calculation which we present now is the QFT version of this calculation. We find that the result of our tree level QFT calculation agrees in the leading order with the result of \cite{Finn:2001qi}. 

From the direct detection of gravitational waves by Virgo and Ligo \cite{TheLIGOScientific:2016src}, mass of the spin-2 graviton is $m_g<1.2\times 10^{-22}\rm{eV}$ which is derived from the experimental upper bound on the dispersion of the gravitational wave event GW150914. 

\subsection{Graviton radiation in massive gravity without vDVZ discontinuity}
In limit $a<R_V$ the Keplarian orbits are also Vainshtein screened similar to FP theory as discussed before and there will be corrections at $\mathcal{O}(n_0)$ in Newtonian potential. Therefore, we consider GR stress-tensor in this case as well. 

 Following the steps described in the Appendix~\ref{appendi}, we compute the rate of energy loss due to the massive graviton radiation as
\begin{eqnarray}
\frac{dE}{dt}&=&\frac{\kappa^2}{8(2\pi)^2}\int \Big[|T_{\mu\nu}(k^\prime)|^2-\frac{1}{2}|T^{\mu}{}_{\mu}(k^\prime)|^2\Big]\delta(\omega-\omega^\prime)\omega^2 \Big(1-\frac{m^2_g}{\omega^2}\Big)^\frac{1}{2}d\omega d\Omega_k\\
&=& \frac{\kappa^2}{8(2\pi)^2}\int \tilde{\Lambda}_{ij,lm}T^{ij*}T^{lm} \delta(\omega-\omega^\prime)\omega^2 \Big(1-\frac{m^2_g}{\omega^2}\Big)^\frac{1}{2}d\omega d\Omega_k,
\label{eq:dedt_deformedFP}
\end{eqnarray}
where
\begin{equation}
\begin{split}
{\tilde{\Lambda}_{ij,lm}}=\Big[\delta_{il}\delta_{jm}-2\left(1-\frac{m_g^2}{\omega^2}\right)\hat{k_j}\hat{k_m}\delta_{il}+\frac{1}{2}\left(1-\frac{m_g^2}{\omega^2}\right)^2\hat{k_i}\hat{k_j}\hat{k_l}\hat{k_m}-\frac{1}{2}\delta_{ij}\delta_{lm}
\\
+\frac{1}{2}\left(1-\frac{m_g^2}{\omega^2}\right)\Big(\delta_{ij}\hat{k_l}\hat{k_m}+\delta_{lm}\hat{k_i}\hat{k_j}\Big)\Big].
\end{split}
\label{eq;Lambda_deformedFP}
\end{equation}
After computation of the angular integration we obtain
\begin{eqnarray}
\frac{dE}{dt}&=& \frac{8G}{5}\int \left[\left\lbrace\frac{5}{2}-\frac{5}{3}\left(1-\frac{m_g^2}{\omega'^2}\right)+\frac{1}{6}\left(1-\frac{m_g^2}{\omega'^2}\right)^2\right\rbrace T^{ij}T^*_{ij}\right.\nonumber\\
&& \left. +\left\lbrace -\frac{5}{4}+\frac{5}{6}\left(1-\frac{m_g^2}{\omega'^2}\right)+\frac{1}{12}\left(1-\frac{m_g^2}{\omega'^2}\right)^2\right\rbrace\vert T^i{}_{i}\vert^2\right]\delta(\omega-\omega^\prime)\omega^2 \Big(1-\frac{m^2_{g}}{\omega^2}\Big)^\frac{1}{2}d\omega. \nonumber\\
\label{eq:a4_deformedFP}
\end{eqnarray}
Finally we get the expression for the rate of energy loss as
\begin{equation}
\begin{split}
\frac{dE}{dt}= \frac{32G}{5} \mu^2 a^4\Omega^6 \sum_{n=1}^{\infty} n^6\sqrt{1-\frac{n_0^2}{n^2}}\left[f(n,e)\left( 1+\frac{4}{3}\frac{n_0^2}{n^2}+\frac{1}{6}\frac{n_0^4}{n^4}\right) -\frac{5J^2_n(ne)}{36n^4}\frac{n_0^2}{n^2}\left(1-\frac{n_0^2}{4n^2}\right)\right].
\end{split}
\label{eq:dedt_FPd}
\end{equation}
To the leading order in $n_0^2$, we can write Eq.\ref{eq:dedt_FPd} as
\begin{equation}
\frac{dE}{dt}\simeq \frac{32G}{5} \mu^2 a^4\Omega^6\Big[\sum_{n=1}^{\infty} n^6 f(n,e)+n^2_0 \sum_{n=1}^{\infty}\Big(\frac{5}{6}n^4 f(n,e)-\frac{5}{36}J^2_n(ne)\Big)\Big]+\mathcal{O}(n^4_0).
\end{equation}
We note that the expression reduces to that of GR in the limit $n_0=0$. Thus there is no vDVZ discontinuity. To the leading order in $n_0^2$ this agrees with the result of the classical calculation of Finn and Sutton \cite{Finn:2001qi}.

\section{ Dvali-Gabadadze-Porrati (DGP) theory}\label{DGP}
The GR theory is a non linear theory which obeys diffeomorphism invariance. However, this symmetry is broken in theories with a massive graviton. In FP theory, if the graviton is expanded around curved spacetime a ghost degree of freedom appears \cite{Boulware:1973my}. To obtain a consistent massive gravity theory free from any ghost, one can go to higher dimension. One such massive gravity theory in higher dimension using a braneworld model framework is the DGP theory \cite{Dvali:2000hr, Dvali:2000rv,Dvali:2000xg,Dvali:2006su}. In the higher dimensions the massless gravity theory has a general covariance symmetry. The number of polarisation states of the spin-2 massless graviton in 5-dimensions is $5$. When the extra dimension compactifes the number of massive graviton degrees of freedom in 4-d remains 5 and there is  no (Boulware Deser) BD ghost. The DGP theory in a cosmological background can account for the cosmological constant \cite{Deffayet:2001pu}. The mass  of gravitons is momentum dependent so that 
one can modify the infrared theory (at cosmological scales) while retaining Newtonian theory at solar system scales. The scalar degree of freedom of the graviton however still contributes to the vDVZ discontinuity which remains a problem for the phenomenological study of the DGP theory of massive gravity \cite{Dvali:2006su}.

In the five dimensional DGP theory, the matter field is localized in a four dimensional brane world which leads to an induced curvature term on the brane. The Planck scales of the five dimensional DGP theory with the four dimensional brane world are denoted by $M_5$ and $M_{pl}$ respectively. 

The action of five dimensional DGP model \cite{Dvali:2000hr,Dvali:2000rv,Dvali:2000xg} with the matter field localized in four dimensional brane world at $y=0$ is
\begin{equation}
\mathcal{S}\supset\int d^4x dy\Big(\frac{M^3_5}{4}\sqrt{-^{(5)}g} {}^{(5)}R+\delta(y)\Big[\sqrt{-g}\frac{M_{pl}^2}{2}R[g]+\mathcal{L}_m(g,\psi_i)\Big]\Big),
\label{j1}
\end{equation}
where $\psi_i$ denotes the matter field with the energy stress tensor $T_{\mu\nu}$ in the brane world.

The resulting modified linearized Einstein equation on the $y=0$ brane is \cite{deRham:2014zqa} 
\begin{equation}
\left(\Box h_{\mu\nu} - \partial_{\mu}\partial_{\nu} h\right) -m_0\sqrt{-\Box}\left(h_{\mu\nu}-h\eta_{\mu\nu}\right)=- \frac{\kappa}{2}T_{\mu\nu}(x),
\end{equation}
where $m_0=\frac{M^3_5}{M^2_{pl}}$, $M_{pl}^2=1/{8\pi G}=4/\kappa^2$. Here, the Fierz-Pauli mass term ($h_{\mu\nu}-h\eta_{\mu\nu}$) appears naturally from the higher dimensional DGP theory. This corresponds to the linearized massive gravity with a scale-dependent effective mass  $m^2_g(\Box)=m_0\sqrt{-\Box}$. 
The propagator is 
\bea
D^{(5)}_{\alpha \beta \mu \nu}(k) &=&  \frac{i}{(-\omega^2 +|\textbf{k}|^2) +m_0 (\omega^2 -|\textbf{k}|^2)^{1/2}} \left (\frac{1}{2} (\eta_{\alpha \mu}  \eta_{\beta \nu} + \eta_{\alpha \nu }\eta_{\beta \mu}) - \frac{1}{3} \eta_{\alpha  \beta}\eta_{\mu \nu} \right).
\label{propDGP}
\eea
The terms in the brackets represent the polarization sum which is identical to that of the
FP theory Eq.\ref{polsumFP}. In the $m_0\rightarrow 0$ limit the DGP propagator does  not go to the massless form Eq.\ref{prop1} and the DGP theory also has the vDVZ discontinuity.

The dispersion relation corresponding to real gravitons in the DGP model is given by the pole  of the propagator Eq.\ref{propDGP},
\begin{equation}
\omega^2= |\textbf{k}|^2 - m_0^2,
\end{equation}
where $|\textbf{k}|$ is the magnitude of the propagation vector. We note that  in the DGP theory the graviton has a tachyonic mass.

Following the same steps of FP theory that we have done in the previous section, we write down the energy loss due to massive graviton radiation in DGP theory. All the relevant expressions in DGP theory differ from those of the FP theory by replacing $m_g^2\rightarrow -m_0^2$ and $\tilde{n}_0^2=m_0^2/\Omega^2=-n_0^2$, i.e. 
\begin{equation}
\frac{dE}{dt}=\frac{\kappa^2}{8(2\pi)^2}\int \Big[|T_{\mu\nu}(k^\prime)|^2-\frac{1}{3}|T^{\mu}{}_{\mu}(k^\prime)|^2\Big]\delta(\omega-\omega^\prime)\omega^2 \Big(1+\frac{m^2_{0}}{\omega^2}\Big)^\frac{1}{2}d\omega d\Omega_k.
\label{j2}
\end{equation}

The components of the stress tensor in $x-y$ plane is given in Eq.\ref{eq:13}. The dispersion relation gives $\hat{k}^i=\frac{k^i}{\omega\sqrt{1+\frac{m^2_0}{\omega^2}}}$. The other components of $T_{\mu\nu}$ can be obtained by using the current conservation relation $k_\mu T^{\mu\nu}=0$ which yields,
\begin{equation}
T_{0j}=-\sqrt{1+\frac{m^2_0}{\omega^2}}\hat{k^i}T_{ij},\hspace{0.5cm} T_{00}=\Big(1+\frac{m^2_0}{\omega^2}\Big)\hat{k^i}\hat{k^j}T_{ij}.
\label{j3}
\end{equation}
Hence, the term in the third bracket of Eq.\ref{j2} can be written in terms of the projection operator $\tilde{\Lambda}_{ij,lm}$ as 
\begin{equation}
\Big[|T_{\mu\nu}(k^\prime)|^2-\frac{1}{3}|T^{\mu}{}_{\mu}(k^\prime)|^2\Big]=\tilde{\Lambda}_{ij,lm}T^{ij*}T^{lm},
\label{j4}
\end{equation}
where
\begin{equation}
{\tilde{\Lambda}_{ij,lm}}=\Big[\delta_{il}\delta_{jm}-2\Big(1+\frac{m^2_0}{\omega^2}\Big)\hat{k_j}\hat{k_m}\delta_{il}+\frac{2}{3}\Big(1+\frac{m^2_0}{\omega^2}\Big)^2\hat{k_i}\hat{k_j}\hat{k_l}\hat{k_m}-\frac{1}{3}\delta_{ij}\delta_{lm}+\frac{1}{3}\Big(1+\frac{m^2_0}{\omega^2}\Big)\Big(\delta_{ij}\hat{k_l}\hat{k_m}+\delta_{lm}\hat{k_i}\hat{k_j}\Big)\Big].
\label{j5}
\end{equation}
In DGP theory, there will be corrections to Newtonian gravitational potential at $\mathcal{O}(n_0)$ in the $a<R_V$ region where Vainshtein screening is active. We can arrive at this from the similar analysis as described in the FP theory. However, in the action \ref{action_decoupling}, there will be non-linearities like \cite{deRham:2012fw},
\begin{equation}
\mathcal{L}_{\rm{int}}\sim \frac{1}{M_{pl}m_g^2}(\partial \phi)^2\Box \phi.
\end{equation}  
At $r=a<<R_V$, deep inside the Vainshtein region, the equation of motion for $\phi$ gives,
\begin{equation}
\frac{\phi}{M_{pl}}\sim m_g\sqrt{\frac{a^3}{R_s}}\frac{R_s}{a} \sim n_0\frac{h}{M_{pl}}
\end{equation}
from balancing $\mathcal{L}_{\rm{int}}\sim \phi^3/(M_{pl}m_g^2r^4)$ against $\phi T/M_{pl} \sim \phi M/(M_{pl}r^3)$, and so the fifth force mediated by the scalar polarisation is only suppressed by $n_0$ relative to the Newtonian force. As before we neglect the correction and consider the GR stress-energy tensor in the calculation of graviton emission rate. 

The final expression of $dE/dt$ for massive DGP theory can be written in the compact form as
\begin{equation}
\begin{split}
\frac{dE}{dt}= \frac{32G}{5} \mu^2 a^4\Omega^6 \sum_{n=1}^{\infty} n^6\sqrt{1+\frac{\tilde{n}_0^2}{n^2}}\left[f(n,e)\left( \frac{19}{18}-\frac{11}{9}\frac{\tilde{n}_0^2}{n^2}+\frac{2}{9}\frac{\tilde{n}_0^4}{n^4}\right) + \frac{5J^2_n(ne)}{108n^4}\left(1+\frac{\tilde{n}_0^2}{n^2}\right)^2\right].
\end{split}
\label{eq:dedt_DGP}
\end{equation}
We can write Eq.\ref{eq:dedt_DGP} to the leading order in $n^2_0$ as
\begin{equation}
\frac{dE}{dt}\simeq \frac{32G}{5} \mu^2 a^4\Omega^6 \Big[\sum_{n=1}^{\infty}\Big(\frac{19}{18}n^6f(n,e)+\frac{5}{108}n^2J^2_n(ne)\Big)-\tilde{n}_0^2\sum_{n=1}^{\infty}\Big(\frac{25}{36}n^4f(n,e)-\frac{25}{216}J^2_n(ne)\Big)\Big]+\mathcal{O}(\tilde{n}_0^4).
\end{equation}

\section{Constraints from observations}\label{Observations}
\begin{table}[h]
\caption{\label{tableI}Summary of the measured orbital parameters and the orbital period derivative values from observation and GR for PSR B1913+16 \cite{Weisberg:2016jye} and PSR J1738+0333 \cite{Freire:2012mg}. The uncertainties in the last digits are quoted in the parenthesis.}
\centering
\begin{tabular}{ |l|c|c|c|c|c| }
 
 \hline
Parameters \hspace{0.01cm} & PSR B1913+16\hspace{0.01cm}&PSR J1738+0333\hspace{0.01cm}\\
 \hline
Pulsar mass $m_1$ (solar masses) &$1.438\pm 0.001$ &$1.46^{+0.06}_{-0.05}$ \\
Companion mass $m_2$ (solar masses)&$1.390\pm 0.001$ & $0.181^{+0.008}_{-0.007}$\\
Eccentricity $e$ &$0.6171340(4)$ &$(3.4\pm 1.1)\times10^{-7}$  \\
Orbital period $P_b$ (d)&$0.322997448918(3)$&$0.3547907398724(13)$\\
Intrinsic $\dot{P_b}(10^{-12}\rm{ss^{-1}})$ &$-2.398\pm 0.004$ &$(-25.9\pm 3.2)\times 10^{-3}$\\
GR $\dot{P_b}(10^{-12}\rm{ss^{-1}})$ &$-2.40263\pm 0.00005$&$-27.7^{+1.5}_{-1.9}\times 10^{-3}$\\
 \hline
\end{tabular}
\end{table}
In this section we calculate the graviton mass from the observation of orbital period decay of the Hulse -Taylor binary system (PSR B1913+16) and a pulsar white-dwarf binary system (PSR J1738+0333). The orbital parameters of the two compact binary systems and the orbital period derivative values from observation and GR are given in TABLE \ref{tableI}. Massless graviton has two states of polarization and the rate of energy loss of the compact binary system due to the emission of massless graviton radiation is given by Eq.~(\ref{eq:app10})
and it agrees with the Peters Mathews formula \cite{Peters:1963ux}.
\begin{figure}
\centering
\includegraphics[width=8.0cm,angle=360]{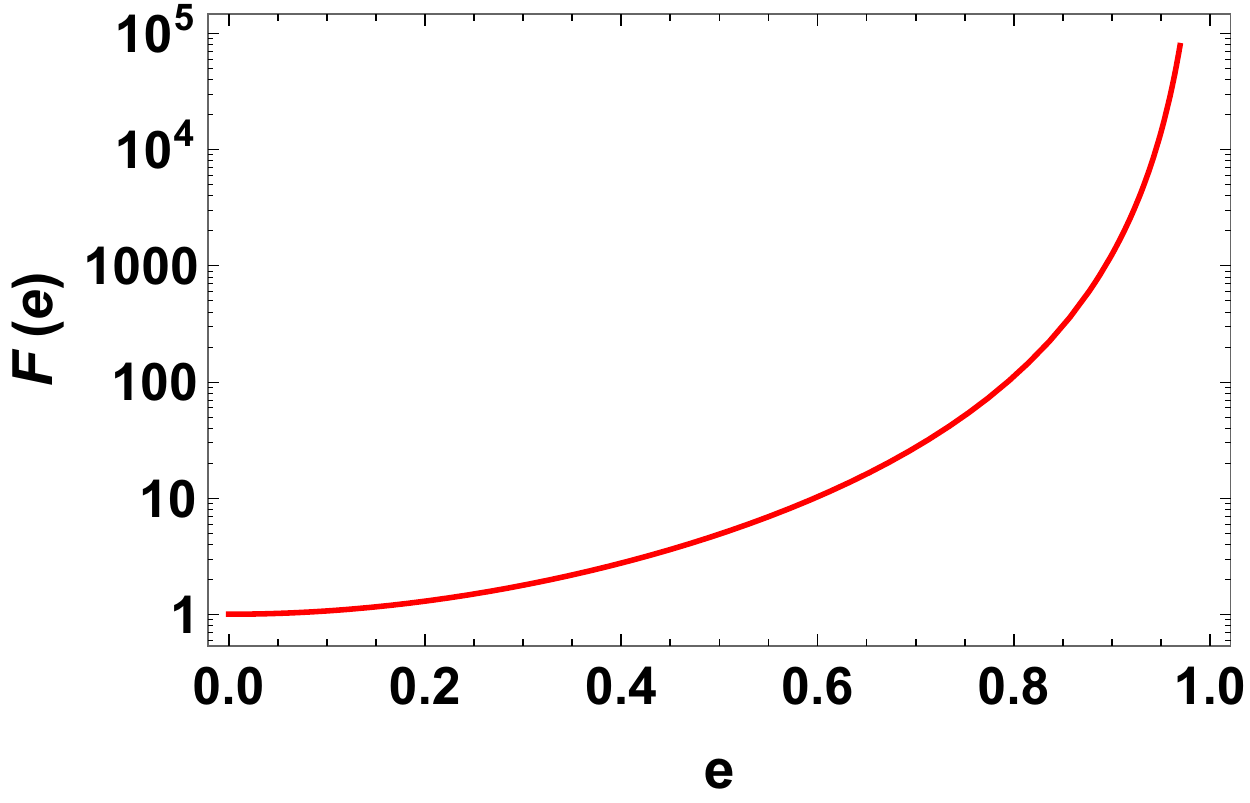}
\caption{Variation of $F(e)$ with the eccentricity.}
\label{fig:feynl}
\end{figure}
The variation of $F(e)=(1-e^2)^{-7/2}\Big(1+\frac{73}{24}e^2+\frac{37}{96}e^4\Big)$ with the eccentricity is shown in FIG.\ref{fig:feynl} which implies that the energy loss due to the GR value is largely enhanced by the eccentricity enhancement factor $F(e)$. Its value is always greater than one for non zero eccentric orbit. Large eccentric binary orbit has strong speed variation as it moves from periastron to apiastron which leads to produce a large amount of radiation in higher harmonics of orbital frequency. In the following, we compare three massive theories of gravity and find limits on the graviton mass for PSR B1913+16 and PSR J1738+0333.

\subsection{Vainshtein radius and limits of linear theory} \label{Vainshtein}
We have used the leading order perturbation of the metric for calculating the graviton emission. In linearised Einstein's gravity the perturbation theory holds as long as $\kappa h_{\mu \nu}\ll1$. This implies that perturbation theory breaks down at radius smaller than $R_s=2 G M$  of the source. If the Fierz-Pauli and no-vDVZ theories are effective field theories describing gravity, with a non-linearly realised diffeomorphism
symmetry, then there will inevitably be interactions below the scale $\Lambda_5\sim (m_g^4M_{pl})^{1/5}$ and these will set the Vainshtein limit of these linearised
theories \cite{Arkani-Hamed:2002bjr}. Therefore, the smallest radius until which the perturbation theory can be applied is Vainshtein radius \cite{Vainshtein:1972sx} given by
\be 
R_V= \left( \frac{R_s}{ m_g^4}\right)^{1/5}\,.
\label{rvFP}
\ee
The Vaishtein radius is much larger than the $R_s$ and perturbative calculations  of the Fierz-Pauli theory are valid in regions with $r>R_V$ away from the source. In our application of binary pulsar radiation, classically the gravitational field is evaluated at the radiation zone such that $R_V < \lambda$ (where $\lambda \sim \pi/\Omega$ is the wavelength of the gravitational waves radiated). In the FP theory this implies that we must have 
\be
\lambda \sim \pi \Omega^{-1} > R_V= \left( \frac{R_s}{ m_g^4}\right)^{1/5}\,.
\ee
We therefore have a lower bound on the graviton mass above which the perturbative calculations is valid given by
\be
m_g > \frac{\Omega^{5/4}}{\pi^{5/4}} ( 2 GM)^{1/4}.
\ee 
Using the numbers as shown in TABLE.\ref{tableI} for PSR B1913+16, we find that the region of  $m_g$ of the Fierz-Pauli theory  where the perturbative calculation is valid is $m_g > 3.06 \times 10^{-22}\, {\rm eV}$. For PSR J1738+0333, we use the Vainshtein limit and obtain the region of graviton mass for the validity of the perturbative calculation    $m_g>2.456\times 10^{-22}\rm{eV}$ for FP theory.


For the DGP theory the Vainshtein radius is given by \cite{Dvali:2000hr,Babichev:2013usa}
\be
R_V= \left( \frac{R_s}{ m_g^2}\right)^{1/3}\,.
\ee
Again we must have $\lambda \sim \pi\Omega^{-1} > R_V$ which gives a lower bound on the graviton mass in the DGP theory above which the perturbative calculation is valid, given by
\be
m_g> \frac{\Omega^{3/2}}{\pi^{3/2}} (2 GM )^{1/2} \,.
\ee
This number is  $7.84 \times 10^{-24} \,{\rm eV}$ for PSR B1913+16 and  $1.406\times 10^{-24}\rm{eV}$
for PSR J1738+0333. 




\subsection{Constraints from observation for FP Theory}
\begin{figure}[!htbp]
\centering
\subfigure[Variation of orbital period loss with graviton mass for PSR B1913+16 in FP theory]{\includegraphics[width=3.0in,angle=360]{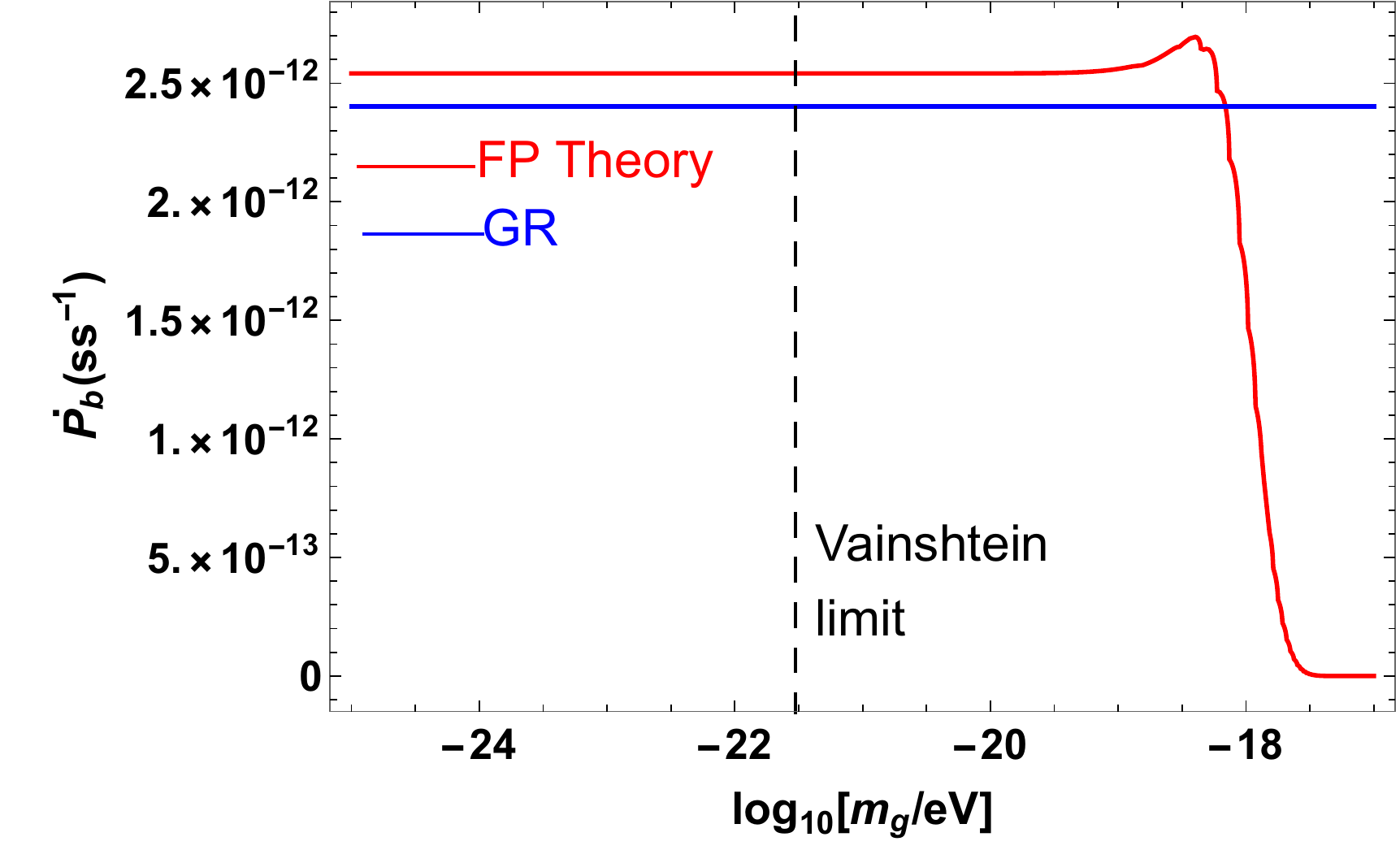}\label{subfig:f1}}
\subfigure[Comparing the theoretical value for the orbital period loss with observation for PSR B1913+16 in FP theory in large $m_g$ limit]{\includegraphics[width=3.0in,angle=360]{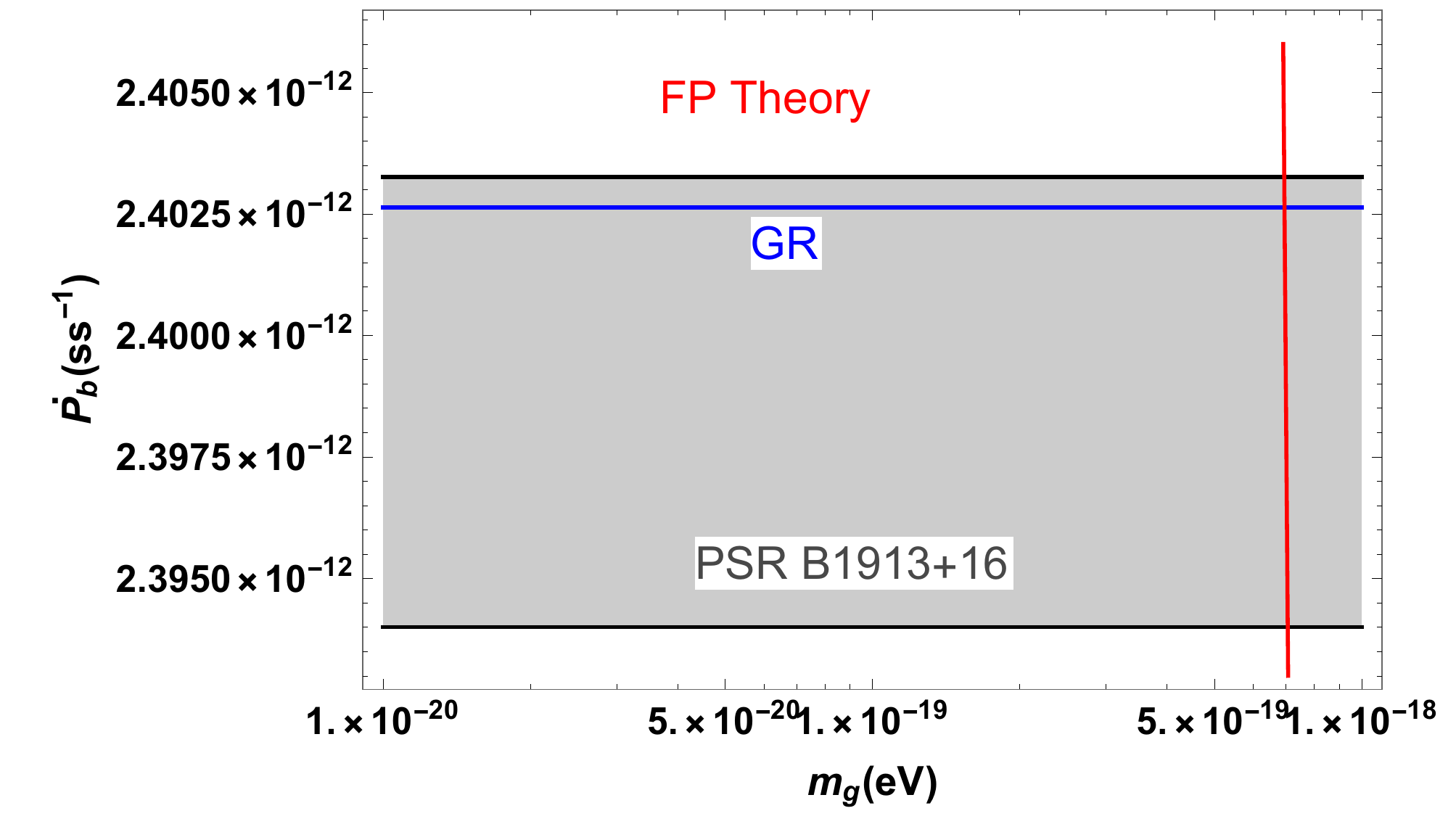}\label{subfig:f2}}
\subfigure[Variation of orbital period loss with graviton mass for PSR J1738+0333 in FP theory]{\includegraphics[width=3.0in,angle=360]{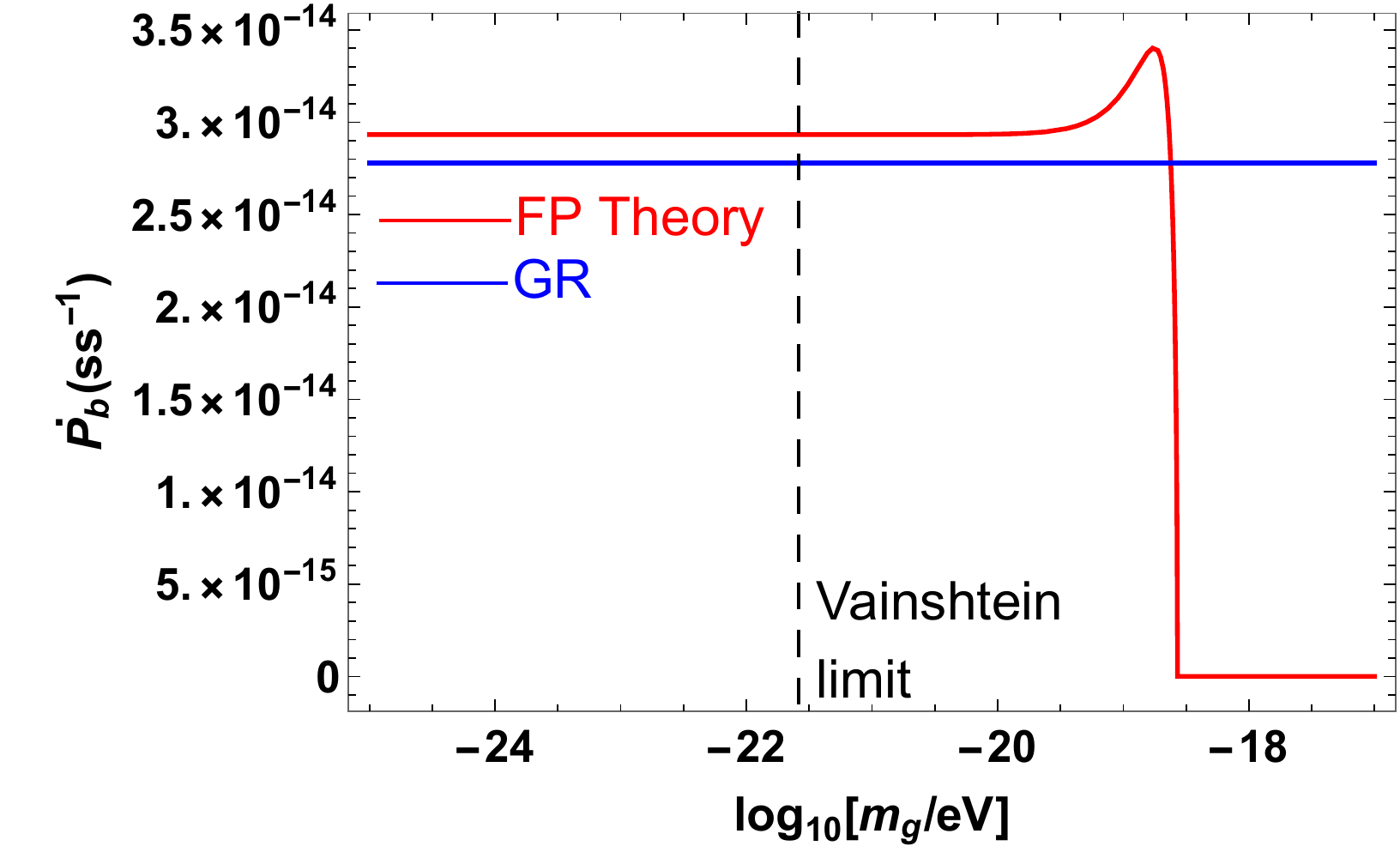}\label{subfig:p1}}
\subfigure[Comparing the theoretical value for the orbital period loss with observation for PSR J1738+0333 in FP theory in large $m_g$ limit]{\includegraphics[width=3.0in,angle=360]{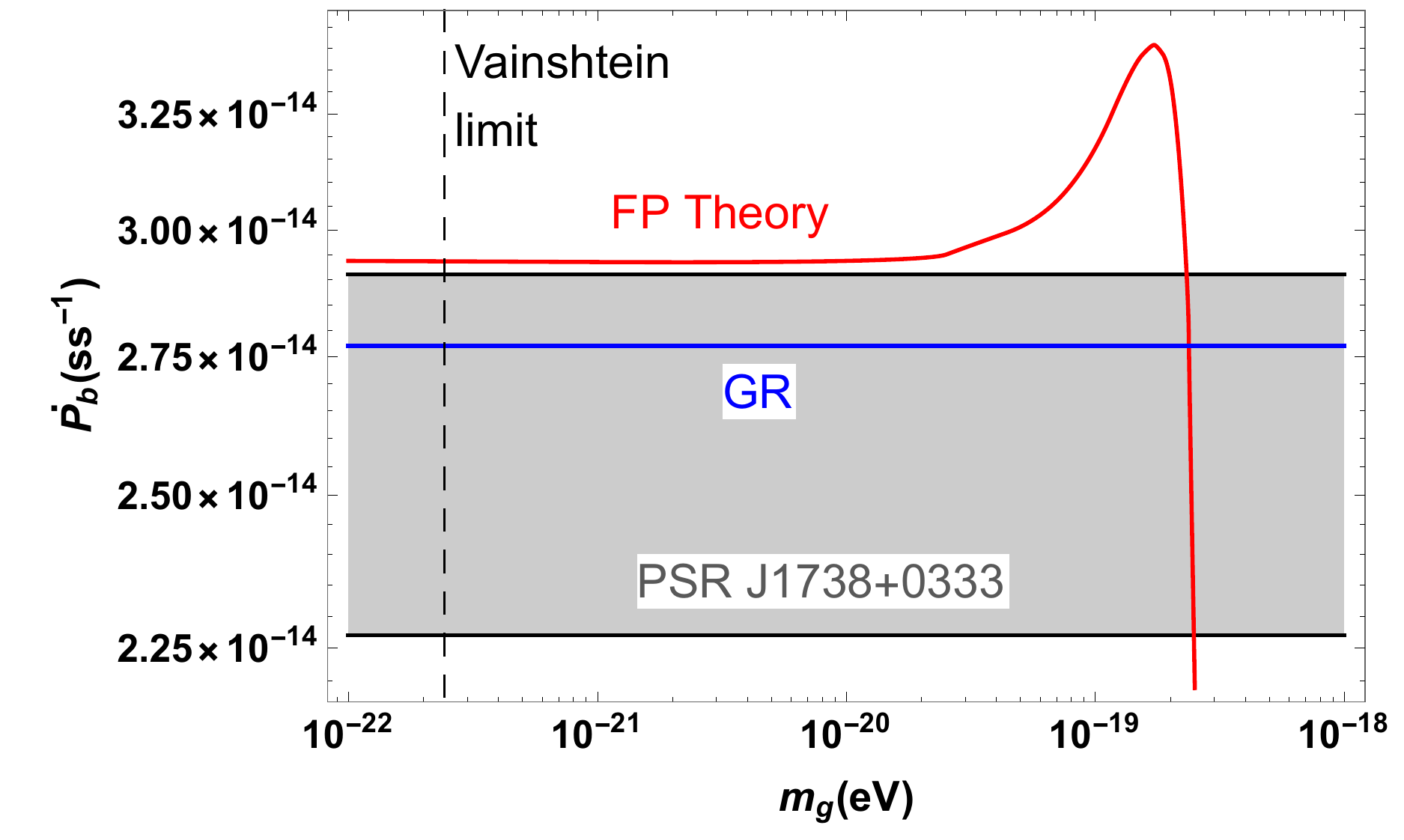}\label{subfig:p2}}
\caption{In the upper panel, we have shown (a) the Variation of orbital period loss with graviton mass and (b) comparing the theoretical value for the orbital period loss with observation for PSR B1913+16 in FP theory. In the lower panel (c) and (d) we have shown the same variation as above for PSR J1738+0333.   }
\label{fig:extremalx}
\end{figure}
The massive graviton has five states of polarization and of these the scalar and the tensor modes couple to the energy momentum tensor and  contribute to the energy loss for the compact binary systems. In the massless limit $m_g\rightarrow 0$ of the FP theory, the extra scalar mode does not decouple and one encounters vDVZ discontinuity. In FIG.\ref{subfig:f1} and FIG.\ref{subfig:f2}, we show the variation of the orbital period loss with the graviton mass for PSR B1913+16 and In FIG.\ref{subfig:p1} and FIG.\ref{subfig:p2} we obtain the same variation for PSR J1738+0333. The dotted lines denote the corresponding Vainshtein limit for the two binary system. The red line denotes the analytical result of orbital period loss in FP theory as obtained above and the blue line denotes the corresponding GR value. The gray band denotes the allowed region of the orbital period loss from observation.

In the region $m_g \sim \Omega$ the energy loss falls with increasing $m_g$ as the phase space of graviton momentum shrinks. There is a region where the theoretical curve goes through the observational band
as shown  FIG.\ref{subfig:f2} and FIG.\ref{subfig:p2} where the variation of orbital period derivative is shown with the observational uncertainty for the two compact binary systems. 


The range of the graviton mass corresponds to $m_g\in (6.88-6.96)\times 10^{-19}\rm{eV}$ (FIG.\ref{subfig:f2}) for PSR B1913+16 and $m_g\in (2.31-2.48)\times 10^{-19}\rm{eV}$ for PSR J1738+0333. There is no common mass range in the overlap region where the red line passes through the gray band for the two binary systems for any value of $m_g$. 

For the FP theory therefore, the limit  on graviton mass from observations PSR B1913+16 together with  PSR J1738+0333 comes from the Vainshtein limit  $m_g>3.06\times 10^{-22}\rm{eV}$.

\subsection{Constraints from observation for DGP Theory}
\begin{figure}[!htbp]
\centering
\subfigure[Variation of orbital period loss with graviton mass for PSR B1913+16 in DGP theory]{\includegraphics[width=3.0in,angle=360]{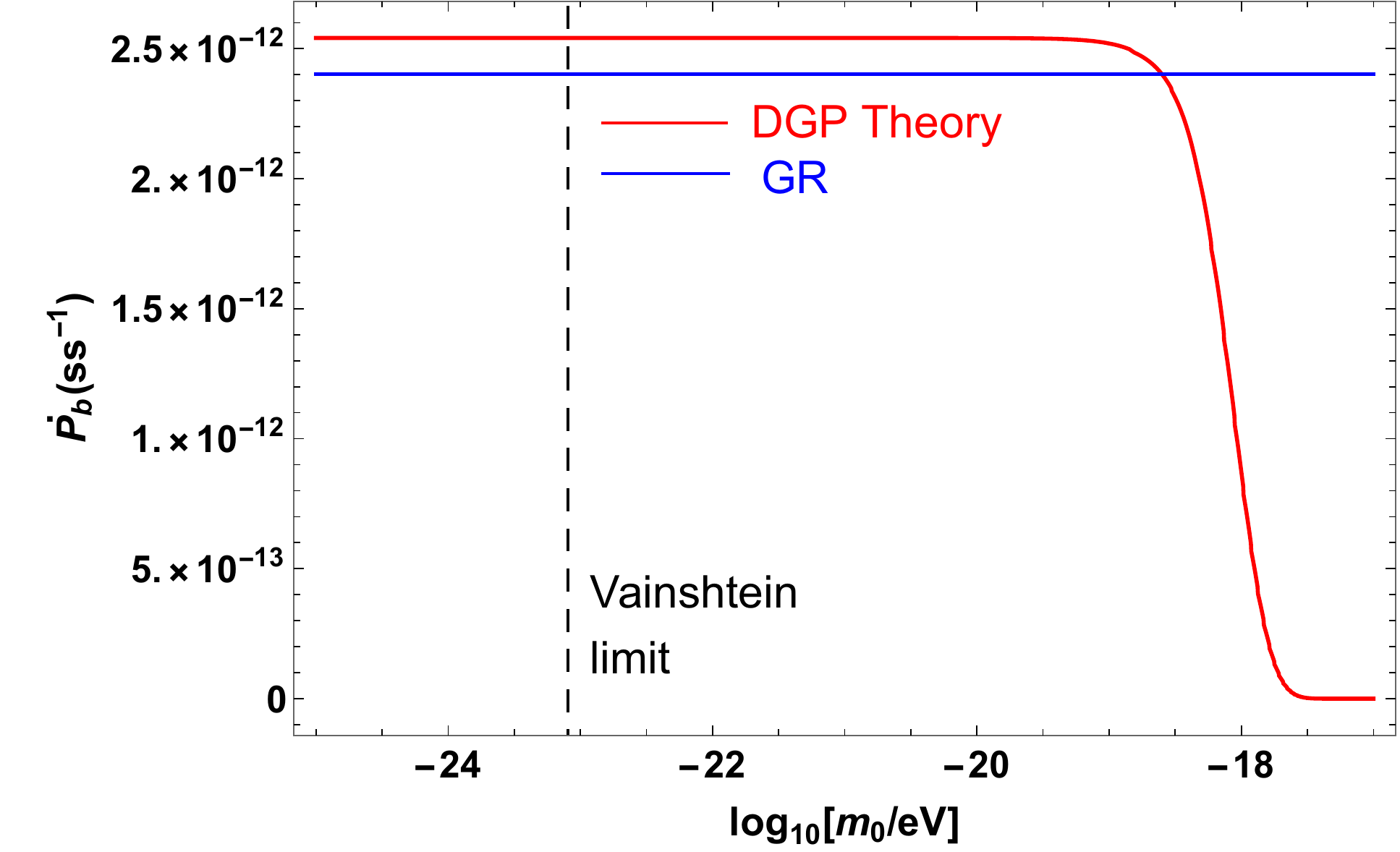}\label{subfig:f3}}
\subfigure[Comparing the theoretical value for the orbital period loss with observation for PSR B1913+16 in DGP theory in large $m_g$ limit]{\includegraphics[width=3.0in,angle=360]{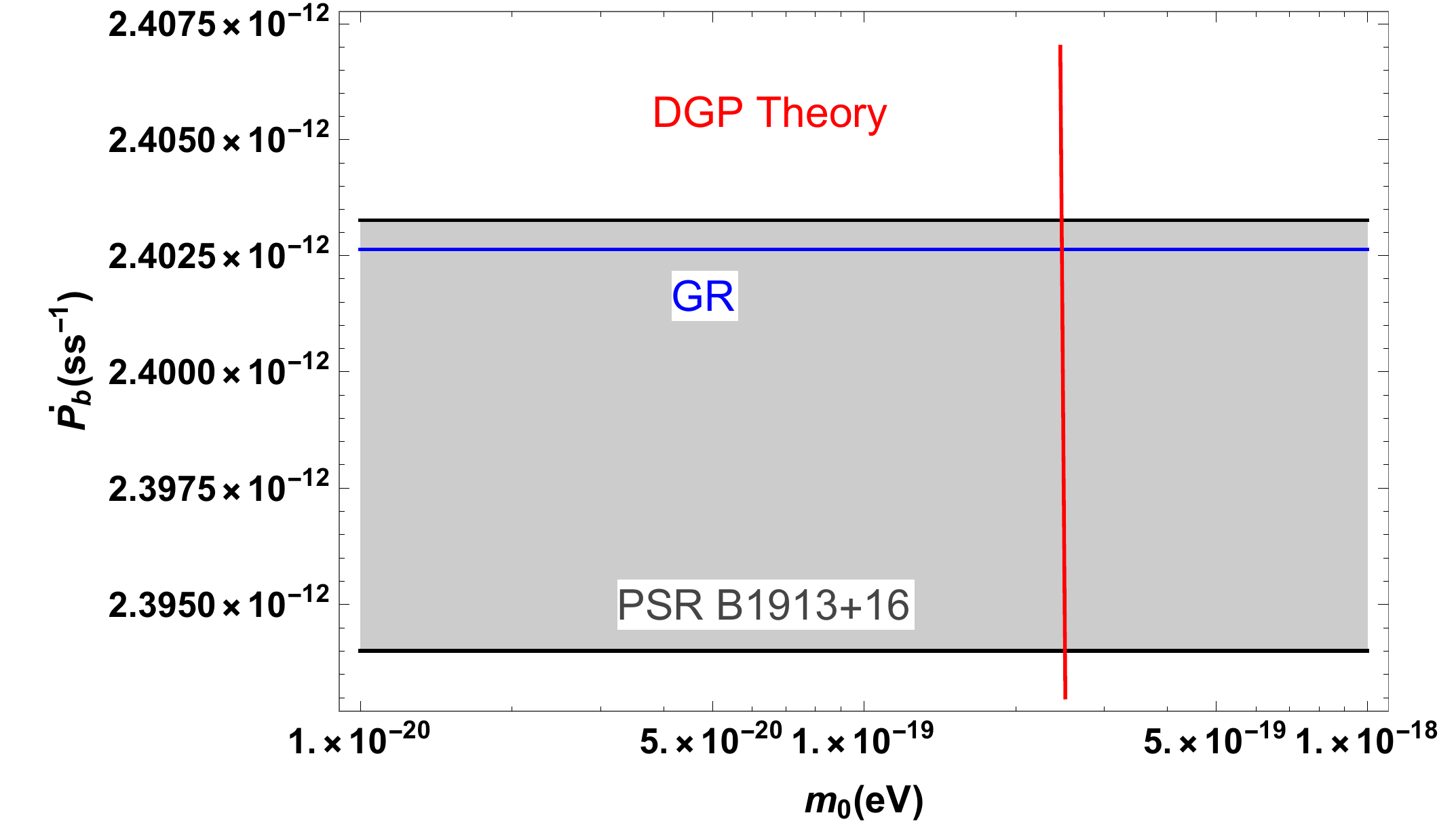}\label{subfig:f4}}
\subfigure[Variation of orbital period loss with graviton mass for PSR J1738+0333 in DGP theory]{\includegraphics[width=3.0in,angle=360]{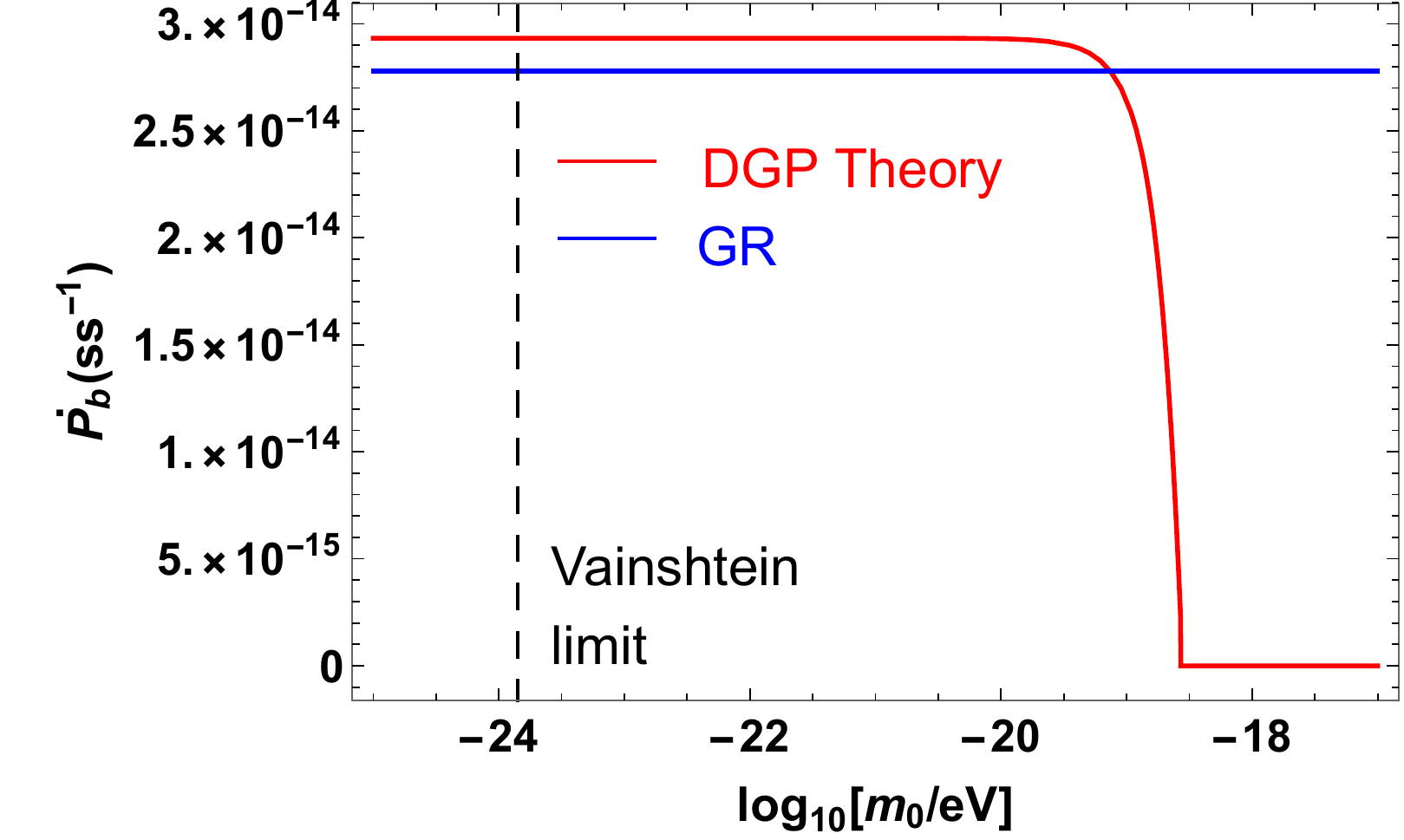}\label{subfig:p3}}
\subfigure[Comparing the theoretical value for the orbital period loss with observation for PSR J1738+0333 in DGP theory in large $m_g$ limit]{\includegraphics[width=3.0in,angle=360]{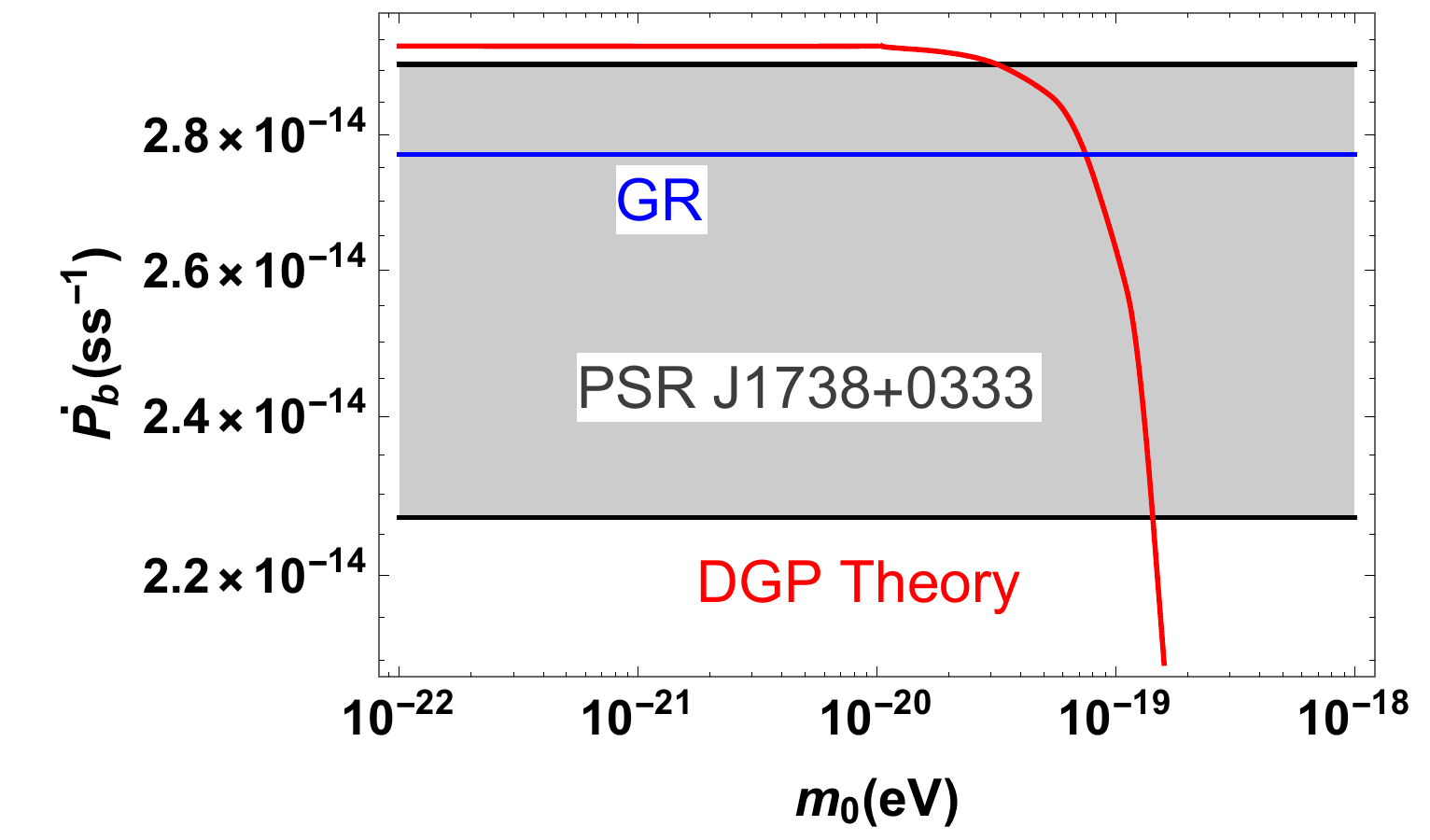}\label{subfig:p4}}
\caption{In the upper panel, we have shown (a) the Variation of orbital period loss with graviton mass and (b) comparing the theoretical value for the orbital period loss with observation for PSR B1913+16 in DGP theory. In the lower panel (c) and (d) we have shown the same variation as above for PSR J1738+0333. }
\label{fig:extremaly}
\end{figure}

In DGP theory, the massless limit $m_0\rightarrow 0$ of the DGP theory does not simply give the massless result and here also one encounters vDVZ discontinuity due to the extra contribution of the scalar gravitons. In FIG.\ref{subfig:f3} and FIG.\ref{subfig:f4}, we show the variation of the orbital period loss with $m_0$ for PSR B1913+16 and in FIG.\ref{subfig:p3} and FIG.\ref{subfig:p4} we obtain the same  for PSR J1738+0333. The dotted lines denote the corresponding Vainshtein limit for the two binary system which are $m_g>7.84\times 10^{-24}\rm{eV}$ for PSR B1913+16 and $m_g>1.406\times 10^{-24}\rm{eV}$ for PSR J1738+0333. 


As in FP theory, in the DGP theory also there is some region where the theoretical prediction crosses the observed band value which corresponds to the graviton mass $m_0\in(2.45-2.47)\times 10^{-19}\rm{eV}$ (FIG.\ref{subfig:f2}) for PSR B1913+16 and $m_0\in(0.31-1.41)\times 10^{-19}\rm{eV}$ for PSR J1738+0333. Since, for DGP theory also, there is no common mass range in the overlap region for the two binary systems, we obtain the graviton mass bound from Vainshtein limit as $m_0 >7.84\times 10^{-24}\rm{eV}$.

\subsection{No vDVZ discontinuity theory}
\begin{figure}[!htbp]
\centering
\subfigure[Variation of orbital period loss with graviton mass for PSR B1913+16 in massive gravity theory without vDVZ discontinuity]{\includegraphics[width=3.0in,angle=360]{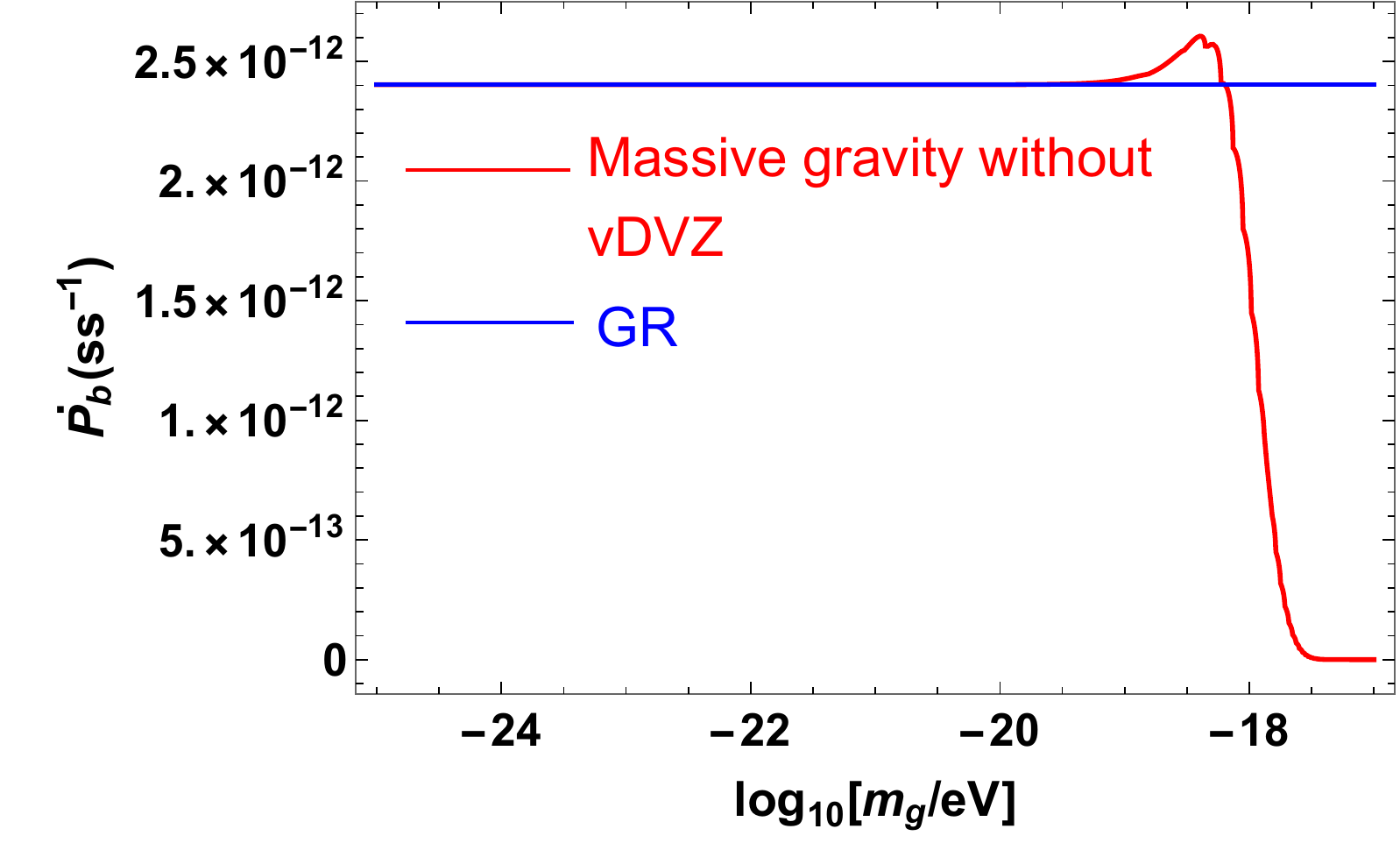}\label{subfig:f5}}
\subfigure[Comparing the theoretical value for the orbital period loss with observation for PSR B1913+16 in massive gravity theory without vDVZ discontinuity for higher graviton mass]{\includegraphics[width=3.0in,angle=360]{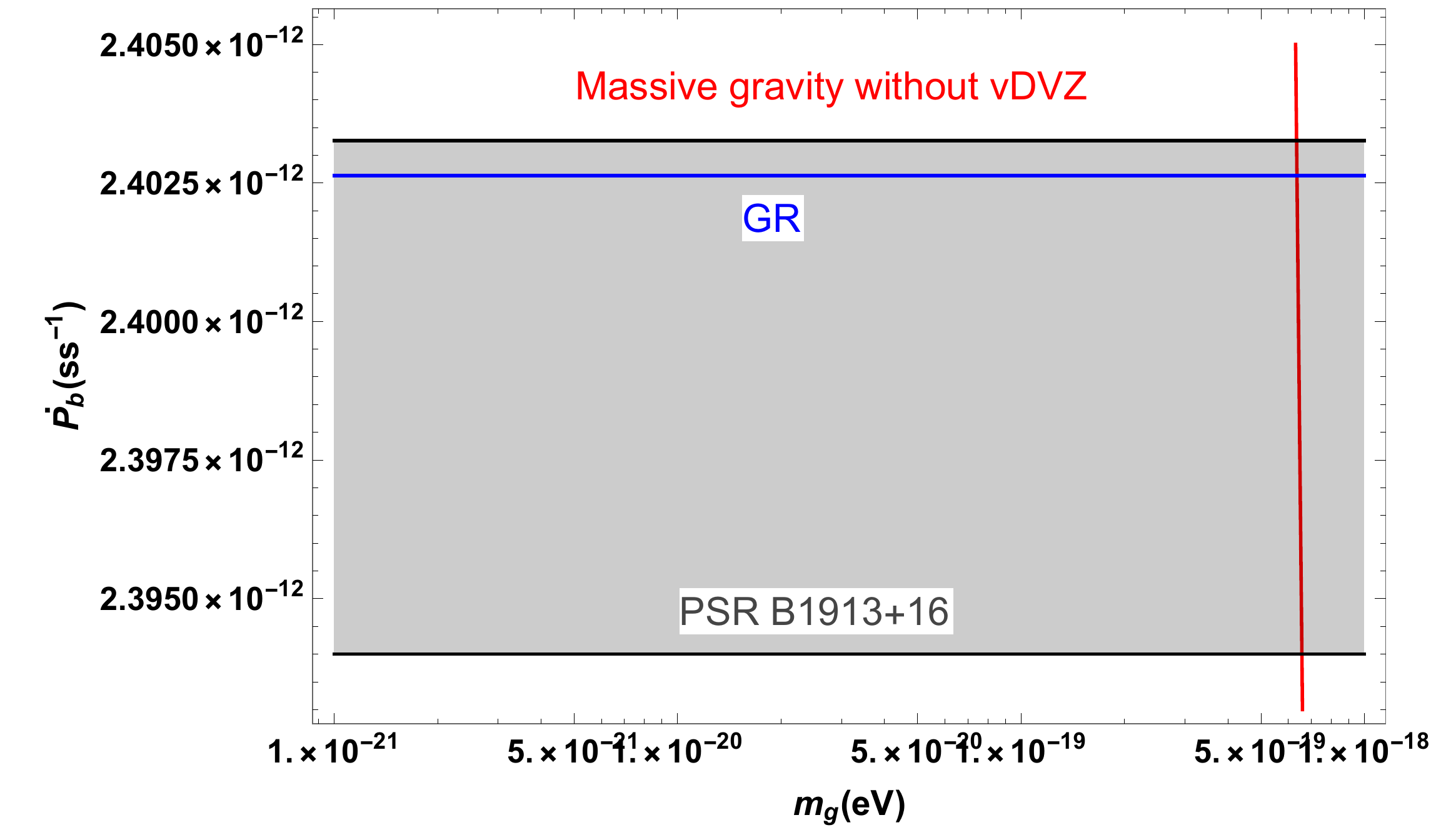}\label{subfig:f6}}
\subfigure[Comparing the theoretical value for the orbital period loss with observation for PSR B1913+16 in massive gravity theory without vDVZ discontinuity for lower graviton mass]
{\includegraphics[width=3.2in,angle=360]{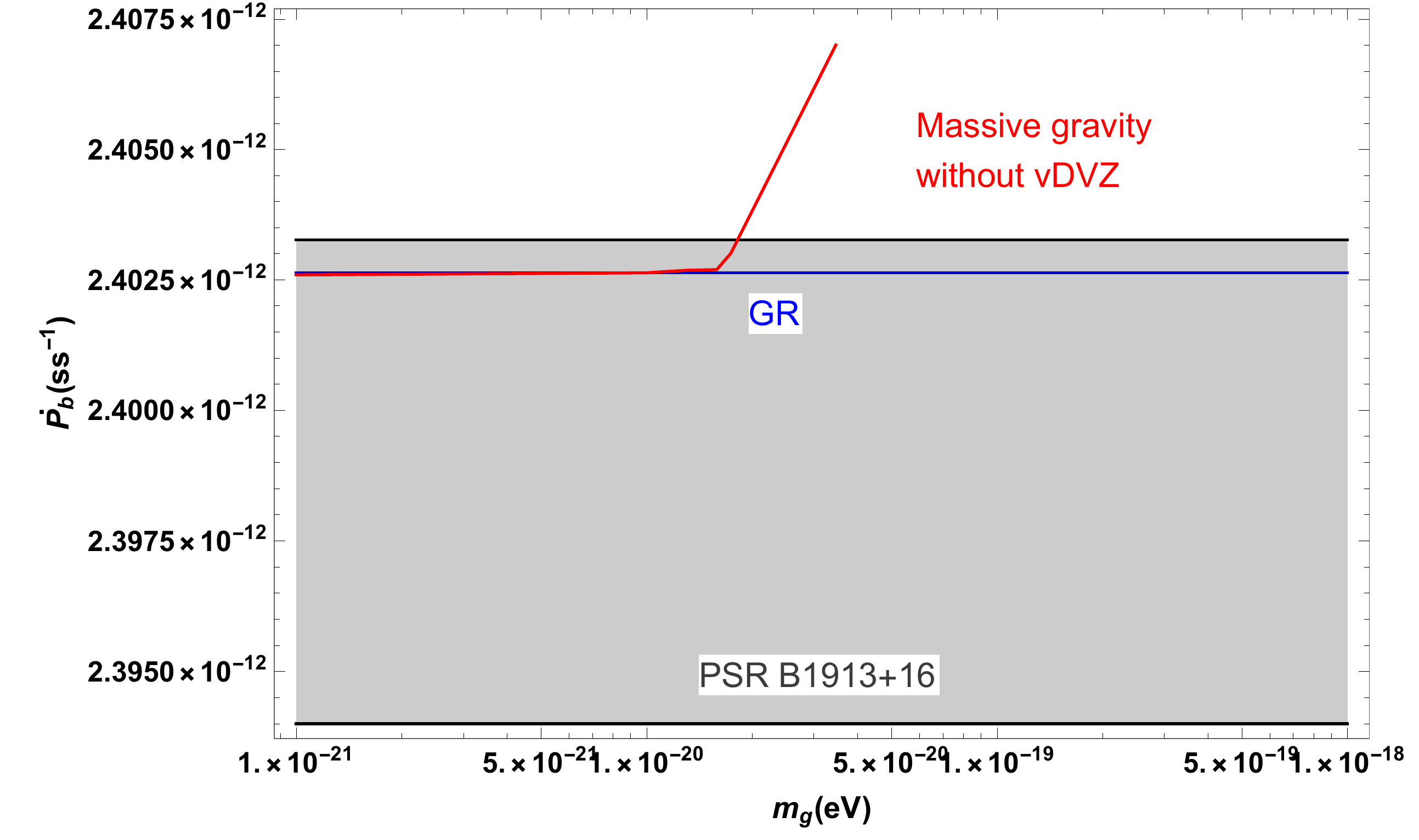}\label{subfig:f7}}
\caption{We have plotted the variation of orbital period loss with graviton mass for PSR B1913+16 in massive gravity theory without vDVZ discontinuity in (a). In (b) and (c) we have compared the theoretical value for the orbital period loss with observation for PSR B1913+16 in massive gravity theory without vDVZ discontinuity for higher graviton mass and lower graviton mass respectively. }
\label{fig:extremalz}
\end{figure}
\begin{figure}[!htbp]
\centering
\subfigure[Variation of orbital period loss with graviton mass for PSR J1738+0333 in massive gravity theory without vDVZ discontinuity]
{\includegraphics[width=3.0in,angle=360]{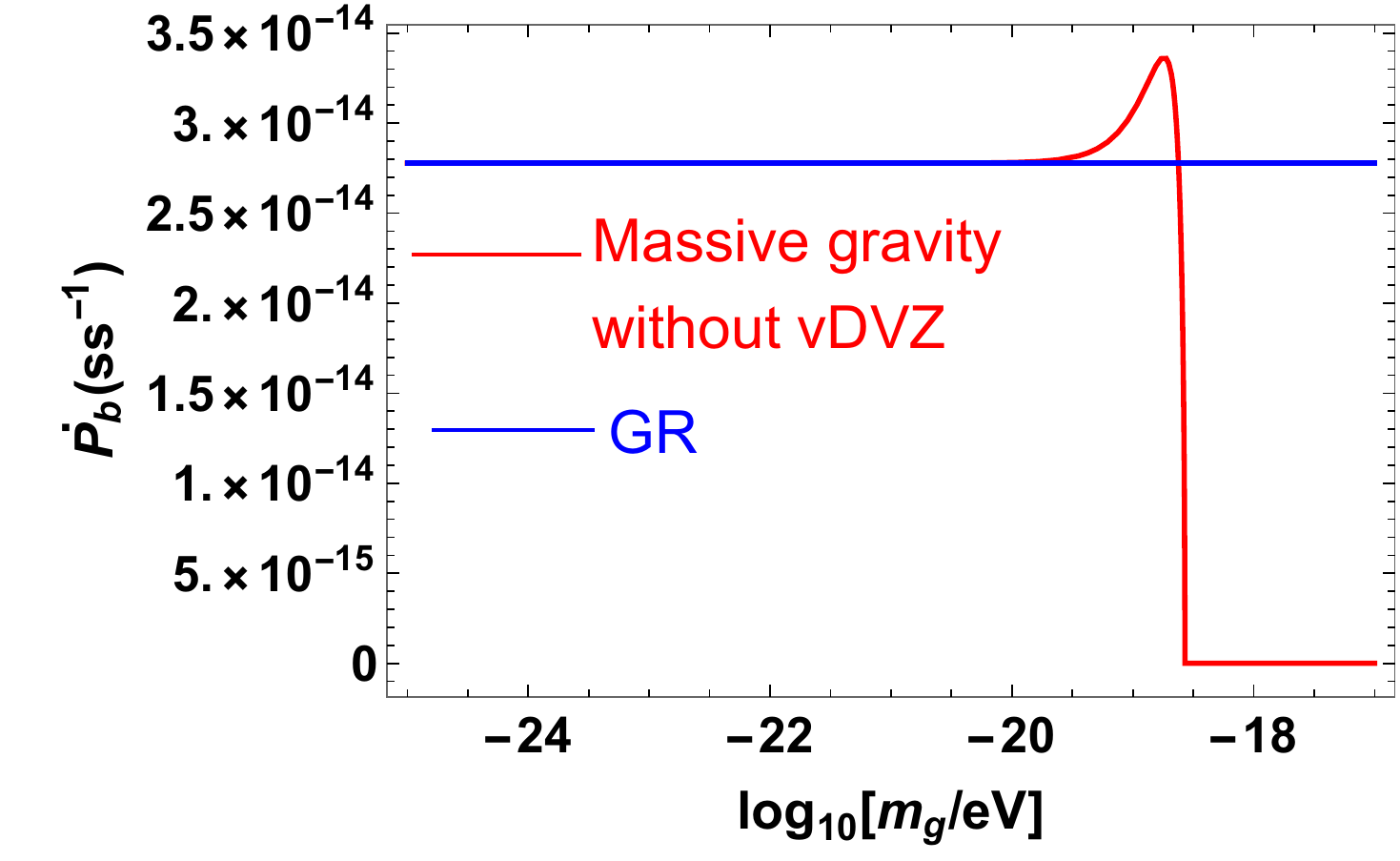}\label{subfig:p5}}
\subfigure[Comparing the theoretical value for the orbital period loss with observation for PSR J1738+0333 in massive gravity theory without vDVZ discontinuity]{\includegraphics[width=3.0in,angle=360]{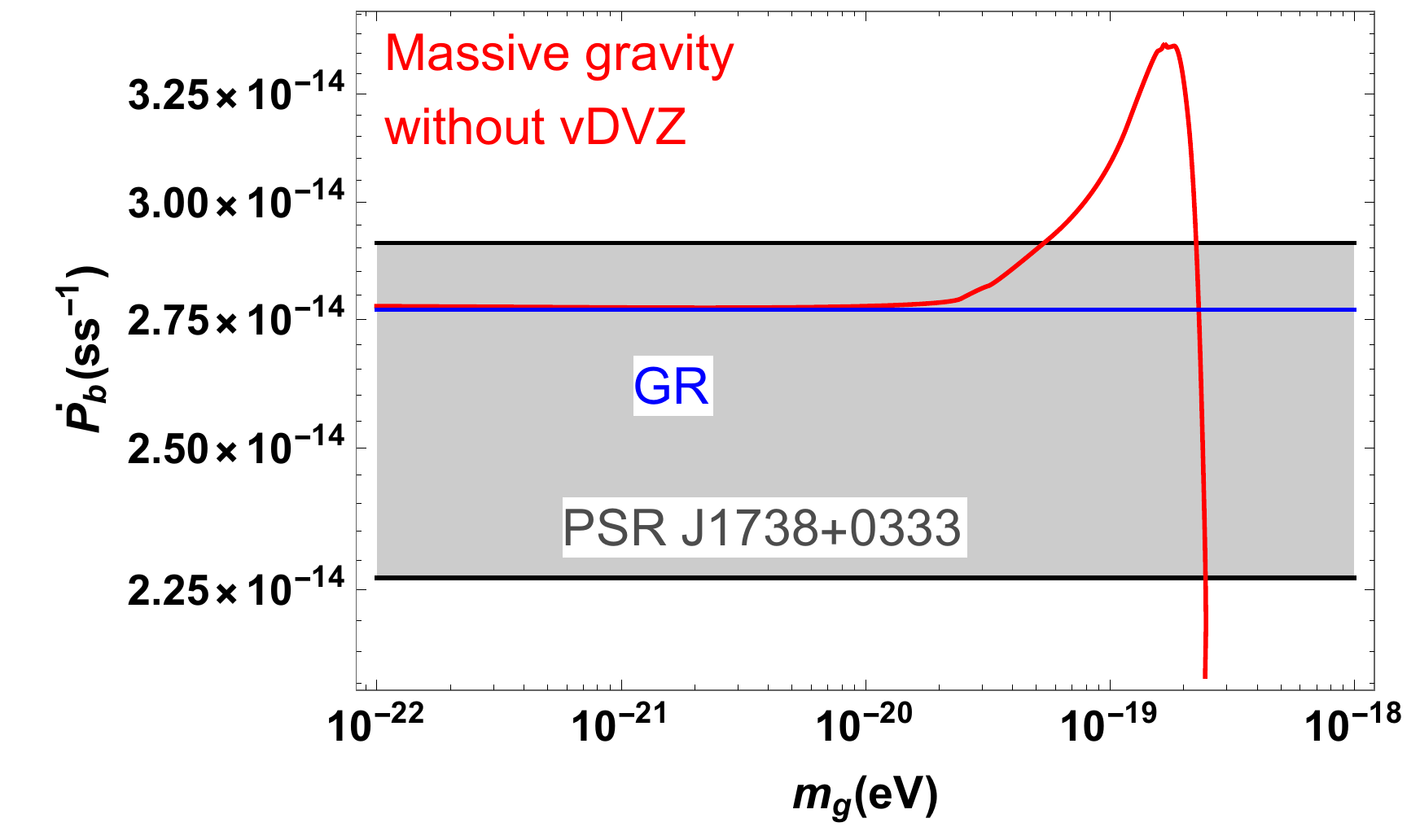}\label{subfig:p6}}
\caption{We have plotted the variation of orbital period loss with graviton mass for PSR J1738+0333 in massive gravity theory without vDVZ discontinuity in (a). In (b) we have compared the theoretical value for the orbital period loss with observation for PSR J1738+0333 in massive gravity theory without vDVZ discontinuity. }
\label{fig:extremalz1}
\end{figure}

The section \ref{FPa} is a special case of massive gravity theory without vDVZ discontinuity at linear order. If one tunes the Fierz-Pauli term $(h_{\mu\nu}h^{\mu\nu}-h^2)$ to $(h_{\mu\nu}h^{\mu\nu}-\frac{1}{2}h^2)$ then at the linear order the ghost term with tachyonic mass cancels the scalar contribution to the propagator. Hence, we are left with the tensor structure of the propagator similar for the massless graviton but having dispersion relation that of massive graviton. Due to this cancellation, there is no vDVZ discontinuity in the $n_0\rightarrow 0$ limit. All our calculations in the paper are at the linear order. However, at the non linear order, there are interactions which will not eliminate the vDVZ discontinuity and the ghost will remain in the theory.

In massive gravity theory without vDVZ discontinuity, the scalar mode is cancelled by the ghost mode. However, there will be Vainshtein radius in the theory similar to FP theory as mentioned in Eq.~(\ref{rvFP}).

In FIG.\ref{fig:extremalz} and FIG.\ref{fig:extremalz1}, we have shown the variation of orbital period loss with the graviton mass for the two compact binary systems and in the low graviton mass limit, the orbital period loss for this theory and massless theory become degenerate. 

There exist two regions where the theoretical prediction  agrees with the observational band. For PSR B1913+16 this  corresponds to the graviton mass $m_g\in (6.32-6.50)\times 10^{-19}\rm{eV}$ and $m_g< 1.81 \times 10^{-20}\rm{eV}$ (FIG.\ref{fig:extremalz}). For PSR J1738+0333, the corresponding graviton mass range are $m_g\in(2.18-2.34)\times 10^{-19}\rm{eV}$ and $m_g < 5.29\times 10^{-20}\rm{eV}$ (FIG.\ref{fig:extremalz1}). Here for the two binary systems we find common graviton mass region where there is agreement with both observations and the bound on graviton mass is $m_g< 1.81 \times 10^{-20}\rm{eV}$. 


All the bounds derived in the paper are at $68\%$ C.L.

\section{Conclusions}\label{Conclusions}
In this paper we put constraints on three massive gravity theories from binary pulsar observations. We show that the bounds on gravitational mass from binary observations are highly model dependent as the predictions for the gravitational luminosity for different graviton mass models  have significant differences. 

In massive gravity theories  like FP and DGP   with an extra propagating scalar, the contribution of the extra scalar to the energy loss  is of the same order as that of the tensor gravitational waves and the region $m_g < \Omega$  is ruled out from binary pulsar observations.  As the graviton mass approaches and becomes larger than $\Omega$ the energy radiated drops with increasing mass. For each binary system there is therefore a range of graviton mass where the theoretical predictions are within observational limits. We found that the allowed ranges of graviton mass from PSR B1913+16 and  PSR J1738+0333 do not have any overlap. Therefore combining observations from the two pulsars we see that no range of graviton mass is consistent with both pulsar observations. In these theories the linear order calculation breaks down below a Vainshtein radius.

The bound on graviton mass from Vainshtein limit is a theoretical bound. Whereas, we describe an independent method of obtaining the mass bound from observation.

In the paper, we have chosen two binary systems PSR B1913+16 and PSR J1738+0333 and compute the orbital period loss for the three massive gravity theories viz, Fierz-Pauli theory, DGP theory and modified Fierz Pauli theory.
Comparing with the observational data, we did not find any overlapping region of graviton mass for FP and DGP theory. For example, in DGP theory, the allowed ranges of mass are $(2.45-2.47)\times 10^{-19}$ eV  for PSR B1913+16 and $(0.31-1.41)\times 10^{-19}$ eV for PSR J1738+0333. So, there is no common allowed mass range valid for both the compact systems and we can not give a universal graviton mass from the observation in DGP theory. Similarly, it is the case for FP theory as well. Therefore, we conclude that for FP and DGP theory the Vainshtein limit puts the stronger bound on the graviton mass.

Before comparing the observational data with our calculation, we cannot tell whether the Vainshtein limit puts stronger limit on graviton mass or not. Although for modified FP theory with no vDVZ discontinuity, we found a common mass region for the two binary systems and put bound on the graviton mass by comparing the observational data with our analytical calculations.

To summarise, observations from PSR B1913+16 and PSR J1738+0333 rule out all value of graviton mass and from the Vainshtein limit we can put the lower bounds $m_g > 3.06 \times 10^{-22
}$ eV for the FP theory and $m_0 > 7.84 \times 10^{-24}$ eV for the DGP theory. 
For the No-vDVZ discontinuity theory the upper bound from combined PSR B1913+16 and PSR J1738+0333 data is $m_g < 1.81 \times 10^{-20} $eV. 
All bounds quoted in the paper are at $68\%$ C.L.

In \cite{Finn:2001qi} the authors used the method of classical multipole expansion of the metric perturbation and kept the term in the expression of the energy loss upto $\mathcal{O}(m^2_g)$. However in our paper, we use the effective field theoretic approach where we treat the graviton as the quantum field and the binary stars as its classical source and we compute the graviton emission rate. The graviton emission is not possible for $\Omega< m_g$ and this is taken care by the factor $(1-m_g^2/\Omega^2)^{1/2}$ in the expression of emission rate. 

In our study the hierarchy of scales is 
\begin{equation}
\frac{a^2}{\lambda^2}\sim \frac{R_s}{a} << \frac{a^2}{R_V^2} <<1 << \frac{R_V}{R_S},
\end{equation}
where $R_s\sim 2M/M_{pl}^2$ and $R_V$ are the usual Schwarzschild and Vainshtein radii around a compact object of mass $M$, $a$ is the orbital separation of the binary, and $\lambda$ is the wavelength of the emitted GW radiation. The condition for graviton emission $\Omega> m_g$ implies that $a< R_V$. This corresponds to a region of space screened  by Vainshtein mechanism. Therefore, we can use the Keplerian orbit in GR in their evaluation of stress-energy tensor $T_{\mu\nu}$. Thus we neglect the corrections in the gravitational potential energy from the screened scalar mode, which are of $\mathcal{O}(n_0)$ for FP and DGP theories. Therefore, our results are approximate and not valid for all orders of $n_0$. These corrections in the Newtonian gravitational potential might change some order unity numerical factors but the order of magnitude of bounds on the graviton mass are expected to be the same as we otained.


In \cite{deRham:2012fw}, the obejective of the paper is different from ours. In this paper, decoupling limit of the DGP theory has been considered, i.e. $M_{pl}\rightarrow \infty$ and $m_g\rightarrow 0$ keeping $m_g^2 M_{pl}$ fixed, where the helicity-2 modes are decoupled from the 0 mode. However, we keep $m_g$ finite. The key difference in our analysis is that we explore the regime $\Omega R_V<< 1$, so that the radiation is described by the linear theory. Where as the paper \cite{deRham:2012fw} use the opposite  $\Omega R_V>> 1$ so that the radiation is Vainshtein screened.
Also, there the authors used the classical multipole expansion method to obtain monopole, dipole, and quadrupole corrections at the leading and subleading orders. Therefore, our method as mentioned earlier is quite different from theirs. 

It should be noted that the upper bound on the graviton mass depends on the length scale of the observation. In fact for DGP theory the mass of the graviton is scale dependant. Naturally, different observation will give different bound on the mass of the graviton. The bounds on the graviton mass mentioned in \cite{deRham:2016nuf} and \cite{Shao:2020fka} are obtained for cubic galileon model which was originally derived from the  decoupling limit of DGP model. However, in our work we have considered the actions for FP, DGP and modified FP theories from the first principle and calculate the energy loss from the binary system using Feynman diagram techniques in the tree level. 
 The bounds on graviton mass that we have obtained are weaker than that for cubic galileon models however our results are comparable with the LIGO bound for direct detection of gravitational waves.
 
Moreover, the calculations for energy loss that we have derived from Feynman diagram techniques are novel and provide interesting results.


There are other massive gravity theories like Lorentz violating gravitational mass \cite{Rubakov:2004eb, Dubovsky:2004sg, Rubakov:2008nh}  and more general Lorentz violating graviton bilinear terms \cite{Kostelecky:2016kfm, Kostelecky:2016uex} which we have not covered in the Lorentz covariant calculation  in this paper. We will address these theories in a separate publication.

The diagramatic method can also be used for computing the wave-form of gravitational waves observed in direct detection experiments like LIGO and VIRGO \cite{TheLIGOScientific:2014jea, TheVirgo:2014hva}. The gravitational wave from extreme mass ratio mergers in massive graviton theories can also constrain the mass of the graviton \cite{Cardoso:2018zhm}. It will be interesting to test massive gravity theory predictions \cite{Will:1997bb,Larson:1999kg} with direct observations and in particular to constrain the scalar and vector modes of gravity from direct detection \cite{TheLIGOScientific:2016src}. 

\section*{Acknowledgements}
The authors are indebted to Vitor Cardoso for mentioning useful constraints. The authors would also like to thank the anonymous referee for useful comments and suggestions.
\appendix
\section{ENERGY LOSS BY MASSLESS GRAVITON RADIATION FROM BINARIES }\label{appendi}
The  action for the graviton field $h_{\mu\nu}$ is obtained by starting with the Einstein-Hilbert action for gravity and matter fields 
\be
S_{EH}= \int d^4 x \sqrt{-g}\left[-\frac{1}{16 \pi G} R + {\cal L}_m \right],
\ee
and expanding the metric $g_{\mu \nu}= \eta_{\mu \nu} + \kappa h_{\mu \nu}$ to the linear order in $h_{\mu \nu}$, where $\kappa= \sqrt{32 \pi G}$ is the gravitational coupling. For consistency the inverse metric $g^{\mu \nu}$ and square root of determinant $\sqrt{-g}$ should be expanded to quadratic order
\bea
g_{\mu \nu}&=& \eta_{\mu \nu} + \kappa h_{\mu \nu}\,,\nonumber\\
g^{\mu \nu}&=& \eta^{\mu \nu} - \kappa h^{\mu \nu} + \kappa^2 h^{\mu \lambda} h^\nu_\lambda +\mathcal{O}(\kappa^3)\,,\nonumber\\
\sqrt{-g}&=& 1+ \frac{\kappa}{2} h +  \frac{\kappa^2}{8} h^2 -  \frac{\kappa^2}{4} h^{\mu \nu}h_{\mu \nu} + \mathcal{O}(\kappa^3),
\eea
where $\eta_{\mu \nu}= {\rm diag}(1,-1,-1,-1)$ is the background Minkowski metric and $h=h^\mu_\mu$. Indices are raised and lowered by $\eta^{\mu \nu}$ and $\eta_{\mu \nu}$ respectively.

The linearised   Einstein-Hilbert action for the graviton field $h_{\mu \nu}$ is then given by
\bea
S_{EH}&=& \int d^4 x \left[ -\frac{1}{2} (\partial_\mu h_{\nu \rho})^2 + \frac{1}{2} (\partial_\mu h)^2 - (\partial_\mu h)(\partial^\nu h^\mu_\nu)+(\partial_\mu h_{\nu \rho} )(\partial^\nu h^{\mu \rho})+ \frac{\kappa}{2} h_{\mu \nu}T^{\mu \nu} \right]\nonumber\\
&\equiv& \int d^4 x \left[ \frac{1}{2} h_{\mu \nu} {\cal E}^{\mu \nu \alpha \beta}h_{\alpha \beta} + \frac{\kappa}{2} h_{\mu \nu }T^{\mu \nu} \right],
\label{EH}
\eea
where the kinetic operator ${\cal E}^{\mu \nu \alpha \beta}$ has the form
\bea
{\cal E}^{\mu \nu \alpha \beta}= \left({\eta^{\mu(\alpha}} {\eta^{\beta)\nu} }-\eta^{\mu \nu}\eta^{\alpha \beta} \right) \Box - \eta^{\mu (\alpha }\partial^{ \beta) } \partial^\nu - \eta^{\nu (\alpha }\partial^{ \beta) } \partial^\mu+ \eta^{\alpha \beta} \partial^\mu \partial^\nu+ \eta^{\mu \nu} \partial^\alpha \partial^\beta ,
\label{kineticOp}
\eea
and indices enclosed by brackets  denote symmetrisation, $A^{(\mu} B^{\nu)}=\frac{1}{2} (A^\mu B^\nu + A^\nu B^\mu)$. The massless graviton propagator  $D^{(0)}_{\mu \nu \alpha \beta}$ is the inverse of the kinetic operator ${\cal E}^{\mu \nu \alpha \beta}$
\be
{\cal E}^{\mu \nu \alpha \beta}D^{(0)}_{\alpha \beta \rho \sigma}(x-y)=  \delta^\mu_{(\rho} \delta^\nu_{\sigma)} \delta^4(x-y)\,.
\label{prope1}
\ee
 The massless graviton action Eq.\ref{EH} has the gauge symmetry $h_{\mu \nu} \rightarrow h_{\mu \nu}- \partial_\mu \xi_\nu -\partial_\nu \xi_\mu$ due to which the operator ${\cal E}^{\mu \nu \alpha \beta}$ cannot be inverted so the propagator cannot be determined from the relation Eq.\ref{prope1}. To invert the kinetic operator we need to choose a gauge. The gauge choice for which the propagator has the simplest form is the de-Dhonder gauge choice in which,
 \be
  \partial^\mu h_{\mu \nu} - \frac{1}{2} \partial_\nu h=0,
 \ee
 where $h={h^\alpha}_{\alpha}$. We can incorporate this gauge choice by adding the  following gauge fixing term to the Lagrangian Eq.\ref{EH},
 \be
 S_{gf}=-  \int d^4x \left( \partial^\mu h_{\mu \nu} - \frac{1}{2} \partial_\nu h\right)^2\,.
 \ee
 The total action with the gauge fixing term turns out to be of the form
 \bea
 S_{EH}+ S_{gf}&=& \int d^4 x  \left(\frac{1}{2} h_{\mu \nu} \Box h^{\mu \nu} -\frac{1}{4} h \Box h + \frac{\kappa}{2} h_{\mu \nu} T^{\mu \nu} \right) \nonumber\\
 &=& \int d^4 x  \left( \frac{1}{2} h_{\mu \nu} {\cal K}^{\mu \nu \alpha \beta} h_{\alpha \beta}  + \frac{\kappa}{2} h_{\mu \nu} T^{\mu \nu} \right),
 \label{Smassless}
 \eea
 where $ {\cal K}^{\mu \nu \alpha \beta}$ is the kinetic operator in the de Donder gauge given by 
 \be
  {\cal K}^{\mu \nu \alpha \beta}= \left(\frac{1}{2} (\eta^{\mu \alpha} \eta^{\nu \beta} +\eta^{\mu \beta}\eta^{\nu \alpha} ) - \frac{1}{2} \eta^{\mu \nu} \eta^{\alpha \beta} \right) \Box\,. 
 \label{Kop} 
 \ee
 The propagator in the de Donder gauge is the inverse of the kinetic operator Eq.\ref{Kop} and is given by 
 \be
{\cal K}^{\mu \nu \alpha \beta}D^{(0)}_{\alpha \beta \rho \sigma}(x-y)=  \delta^\mu_{(\rho} \delta^\mu_{\sigma)}\delta^4(x-y)\,.
\label{Prop1}
\ee
This relation can be used to solve for $D^{(0)}_{\alpha \beta \rho \sigma}$ which in the momentum space  ($\partial_\mu = i k_\mu$)   is then given
\be
D^{(0)}_{\mu \nu \alpha \beta}(k)=  \frac{1}{-k^2} \left( \frac{1}{2}(\eta_{\mu\alpha}\eta_{\nu\beta}+\eta_{\mu\beta}\eta_{\nu\alpha})- \frac{1}{2}\eta_{\mu\nu}\eta_{\alpha\beta} \right)\,.
\label{prop1}
\ee
We treat the graviton as a quantum field by expanding it in terms of creation and annihilation operators,
\be
\hat {h}_{\mu \nu}(x)= \sum_\lambda \int \frac{d^3 k}{(2 \pi)^3} \frac{1}{\sqrt{2 \omega_k}} \left[ \epsilon^\lambda_{\mu \nu}(k) a_\lambda(k) e^{-i k \cdot x} +  \epsilon^{*\lambda}_{\mu \nu}(k) a^\dagger_\lambda(k) e^{i k \cdot x}  \right]\,.
\label{hhat}
\ee
Here $\epsilon^\lambda_{\mu \nu}(k)$ are the polarization tensors which obey the orthogonality relation
\be 
 \epsilon^\lambda_{\mu \nu}(k) {\epsilon^{*\lambda^\prime}}^{\mu \nu}(k)= \delta_{\lambda \lambda^\prime},
\ee
while $a_\lambda(k)$ and $ a^\dagger_\lambda(k)$ are graviton annihilation and creation operators which obey the canonical commutation relations
\be
\left[a_\lambda(k), a^\dagger_{\lambda^\prime}(k^\prime)\right]= \delta^4(k-k^\prime) \delta_{\lambda \lambda^\prime}\,.
\ee

The Feynman propagator of gravitons is defined as the time ordered two point function
\be
D^{(0)}_{\mu \nu \alpha \beta}(x-y) \equiv \langle 0 | T (\hat {h}_{\mu \nu}(x) \hat {h}_{\alpha \beta}(y))|0\rangle,
\ee
which may be evaluated using Eq.\ref{hhat} to give
\be
D^{(0)}_{\mu \nu \alpha \beta}(x-y)=\int \frac{d^4 k}{(2 \pi)^4}\frac{1}{-k^2 +i \epsilon} e^{i k(x-y)} 
\sum_{\lambda} \epsilon_{\mu\nu}^\lambda(k)\epsilon_{ \alpha\beta}^{*\lambda}(k).
\label{prop2}
\ee
Comparing Eq.\ref{prop1} and Eq.\ref{prop2} we have the expression for the 
 polarization sum of massless spin-2 gravitons  
\begin{equation}
\sum_{\lambda=1}^2 \epsilon_{\mu\nu}^\lambda(k)\epsilon_{ \alpha\beta}^{*\lambda}(k)= \frac{1}{2}(\eta_{\mu\alpha}\eta_{\nu\beta}+\eta_{\mu\beta}\eta_{\nu\alpha})- \frac{1}{2}\eta_{\mu\nu}\eta_{\alpha\beta}  .
\label{app3}
\end{equation}
This will be used in the computation of massles graviton radiation from classical sources.

 We now calculate the energy loss due to the radiation of massless graviton from compact binary systems \cite{Mohanty:1994yi, Mohanty:2020pfa} by evaluating the Feynman diagram shown in FIG.\ref{fig:feyn}. We treat the current $T_{\mu \nu}$ of the binary stars as classical source and the gravitons as quantum fields. From the interaction Lagrangian Eq.\ref{Smassless} we see that the interaction vertex is $\frac{1}{2}\kappa h^{\mu\nu}T_{\mu\nu}$, therefore we can write the emission rate of massless gravitons with polarisation tensor $\epsilon_\lambda^{\mu \nu}(k^\prime)$ from the classical source $T_{\mu \nu}(k)$ as
\begin{equation}
d\Gamma= \frac{\kappa^2}{4}\sum_{\lambda=1}^2|T_{\mu\nu}(k^\prime)\epsilon^{\mu\nu}_\lambda (k)|^2 2\pi \delta(\omega-\omega^\prime)\frac{d^3k}{(2\pi)^3}\frac{1}{2\omega}.
\label{app1}
\end{equation}   
Expanding the modulus squared in Eq.\ref{app1}, we can write 
\begin{equation}
d\Gamma=\frac{\kappa^2}{8(2\pi)^2}\sum_{\lambda=1}^2\Big(T_{\mu\nu}(k^\prime)T^*_{\alpha\beta}(k^\prime)\epsilon^{\mu\nu}_\lambda(k)\epsilon^{*\alpha\beta}_\lambda(k)\Big)\frac{d^3k}{\omega}\delta(\omega-\omega^\prime).
\label{app2}
\end{equation}
Using the polarization sum of massless spin-2 gravitons from 
Eq.\ref{app3}, we can write the emission rate as
\bea
d\Gamma&=&\frac{\kappa^2}{8(2\pi)^2}\int \Big[T_{\mu\nu}(k^\prime)T^*_{\alpha\beta}(k^\prime)\Big]\Big[\frac{1}{2}(\eta^{\mu\alpha}\eta^{\nu\beta}+\eta^{\mu\beta}\eta^{\nu\alpha}-\eta^{\mu\nu}\eta^{\alpha\beta})\Big]\frac{d^3k}{\omega}\delta(\omega-\omega^\prime)\,.\nonumber\\
&=&\frac{\kappa^2}{8(2\pi)^2}\int \Big[|T_{\mu\nu}(k^\prime)|^2-\frac{1}{2}|T^{\mu}{}_{\mu}(k^\prime)|^2\Big]\delta(\omega-\omega^\prime)\omega d\omega d\Omega_k,
\label{app4}
\eea
where we use $d^3k=k^2dkd\Omega$. Thus, the rate of energy loss due to massless graviton radiation becomes
\begin{equation}
\frac{dE}{dt}=\frac{\kappa^2}{8(2\pi)^2}\int \Big[|T_{\mu\nu}(k^\prime)|^2-\frac{1}{2}|T^{\mu}{}_{\mu}(k^\prime)|^2\Big]\delta(\omega-\omega^\prime)\omega^2d\omega d\Omega_k.
\label{app5}
\end{equation}
Using the conserved current relation $k_\mu T^{\mu\nu}=0$, we can write the $T^{00}$ and $T^{i0}$ components of the stress tensor in terms of the $T^{ij}$ components,
\begin{equation}
T_{0j}=-\hat{k^i}T_{ij},\hspace{0.5cm} T_{00}=\hat{k^i}\hat{k^j}T_{ij}.
\label{app6}
\end{equation}
Therefore, we can write
\begin{equation}
\Big[|T_{\mu\nu}(k^\prime)|^2-\frac{1}{2}|T^{\mu}{}_{\mu}(k^\prime)|^2\Big]={\Lambda^0_{ij,lm}}T^{ij*}T^{lm},
\label{app7}
\end{equation}
where,
\begin{equation}
{\Lambda^0_{ij,lm}}=\Big[\delta_{il}\delta_{jm}-2\hat{k_j}\hat{k_m}\delta_{il}+\frac{1}{2}\hat{k_i}\hat{k_j}\hat{k_l}\hat{k_m}-\frac{1}{2}\delta_{ij}\delta_{lm}+\frac{1}{2}\Big(\delta_{ij}\hat{k_l}\hat{k_m}+\delta_{lm}\hat{k_i}\hat{k_j}\Big)\Big].
\label{app8}
\end{equation}
We do the angular integrals
\begin{equation}
\int d\Omega_k \Lambda^0_{ij,lm}T^{ij*}(\omega^\prime)T^{lm}({\omega^\prime})=\frac{8\pi}{5}\Big(T_{ij}(\omega^\prime)T^*_{ji}(\omega^\prime)-\frac{1}{3}|T^{i}{}_{i}(\omega^\prime)|^2\Big),
\label{app9}
\end{equation}
using the relations
\begin{equation}
\int d\Omega \hat{k^i}\hat{k^j}=\frac{4\pi}{3}\delta_{ij}, \hspace{0.5cm} \int d\Omega \hat{k^i}\hat{k^j}\hat{k^l}\hat{k^m}=\frac{4\pi}{15}(\delta_{ij}\delta_{lm}+\delta_{il}\delta_{jm}+\delta_{im}\delta_{jl}).
\label{dOmegak}
\end{equation}
The stress tensor or the current density for this compact binary system is
\begin{equation}
T_{\mu\nu}(x^\prime)=\mu \delta^3(\textbf{x}^\prime-\textbf{x}(t))U_\mu U_\nu,
\label{eq:7}
\end{equation}
where $\mu=\frac{m_1m_2}{m_1+m_2}$ is the reduced mass of the binary system and $m_1$ and $m_2$ are the masses of the two stars in the binary system. $U_\mu=(1,\dot{x},\dot{y},0)$ is the non relativistic four velocity of the reduced mass of the binary system in the $x-y$ plane of the Keplerian orbit.

This stress energy tensor only corresponds to the matter fields but not the effective stress-energy tensor which, in general, is $T_{\mu\nu}=T^{matter}_{\mu\nu}+ T^{GW}_{\mu\nu}$, $ T^{matter}_{\mu\nu}$ is the usual stress-energy tensor for matter fields and $ T^{GW}_{\mu\nu}$ corresponds to the energy content of the gravitational waves. The expression for the $ T^{GW}_{\mu\nu}$ is
\begin{equation}
T_{\mu\nu}^{GW}= \langle h_{\alpha\beta,\mu}h^{\alpha\beta}{}_{,\nu} -\frac{1}{2}h_{,\mu}h_{,\nu} \rangle.
\end{equation}
Now at the tree-level, from the equation of motion for $h_{\alpha\beta}$, we can write
\begin{equation}
h_{\alpha\beta}\sim \frac{1}{M_{pl}}(\Box -m^2_g)^{-1}T^{matter}_{\alpha\beta} .
\end{equation}
Therefore,
\begin{equation}
T_{\mu\nu}^{GW}(k_{\alpha})\sim \frac{1}{M_{pl}^2}\left((T^{matter}_{\alpha\beta})^2-\frac{(T^{matter})^2}{2}\right)\left(\frac{k_{\mu}k_{\nu}}{(k^{\alpha}k_{\alpha}-m^2_g)^2}\right) .
\end{equation}
Thus in the radiation zone, i.e. far from the source, $T^{GW}_{\mu\nu}$ is suppressed by the factor of $1/M_{pl}^2$ in comparison with the part ($T_{\mu\nu}^{matter}$) from the matter field. Therefore, for gravitational radiation from compact binaries, $T_{\mu\nu}\simeq T^{matter}_{\mu\nu}$.

We can write the Keplerian orbit in the parametric form as
\begin{equation}
x=a(\cos\xi-e), \hspace{0.4cm} y=a\sqrt{(1-e^2)}\sin\xi, \hspace{0.4cm} \Omega t=\xi-e\sin\xi,
\label{eq:8}
\end{equation}
where $a$ and $e$ are the semi-major axis and eccentricity of the elliptic orbit respectively. Since the angular velocity of an eccentric orbit is not constant, we can write the Fourier transform of the current density in terms of the $n$ harmonics of the fundamental frequency $\Omega=\Big[G\frac{(m_1+m_2)}{a^3}\Big]^\frac{1}{2}$. Using Eq.\ref{eq:8}, we can write the Fourier transforms of the velocity components in the Kepler orbit as
\begin{equation}
\dot{x}_n=\frac{1}{T}\int ^T_0 e^{in\Omega t}\dot{x}dt=-ia\Omega J^\prime_n(ne),
\label{eq:9}
\end{equation}  
and
\begin{equation}
\dot{y}_n=\frac{1}{T}\int ^T_0 e^{in\Omega t}\dot{y}dt=\frac{a\sqrt{(1-e^2)}}{e}\Omega J_n(ne),
\label{eq:10}
\end{equation}
where we have used $T=2\pi/\Omega$ and the Bessel function identity $J_n(z)=\frac{1}{2\pi}\int^{2\pi}_0 e^{i(n\xi-z\sin\xi)}d\xi$. The prime over the Bessel function denotes the derivative with respect to the argument. Hence the Fourier transforms of the orbital coordinates become
\begin{equation}
x_n=\frac{\dot{x}_n}{-i\Omega n}=\frac{a}{n}J^\prime_n (ne),\hspace{0.4cm} y_n=\frac{\dot{y}_n}{-i\Omega n}=\frac{ia\sqrt{1-e^2}}{ne}J_n(ne).
\label{eq:11}
\end{equation}
Now we will calculate the Fourier transforms of different components of the stress tensor with $\omega^\prime=n\Omega$ as below. 
Thus,
\begin{eqnarray}
T_{ij}(\textbf{k}^\prime,\omega^\prime)&=&\frac{1}{T}\int_0^T\int T_{ij}(\textbf{x},t) e^{-i(\textbf{k}^\prime\cdot \textbf{x}-\omega^\prime t)}d^3x dt\nonumber\\
&=& \int T_{ij}(\textbf{x},\omega^{\prime}) e^{-i\textbf{k}^\prime\cdot \textbf{x}}d^3x.
\label{eq:12} 
\end{eqnarray}
Expanding $e^{-i\textbf{k}^\prime.\textbf{x}}\approx1-i\textbf{k}^\prime.\textbf{x}-\cdots$ and retaining the leading order term $\textbf{k}^\prime.\textbf{x}\sim \Omega a\ll 1$ for binary orbit, we can write Eq.\ref{eq:12} as
\begin{equation}
T_{ij}(\textbf{k}^\prime,\omega^\prime)\simeq T_{ij}(\omega^\prime)= \int T_{ij}(\textbf{x},\omega^{\prime}) d^3x.
\end{equation}
From conservation of the stress-energy tensor, i.e. $\partial_{\mu}T^{\mu\nu}(\textbf{x},t)=0$, we get 
\begin{equation}
\partial^i\partial^j T_{ij}(\textbf{x},\omega^{\prime})= -\omega^{\prime}{}^{2}T_{00}(\textbf{x},\omega^{\prime}).
\label{eq:cons_Tmunu}
\end{equation}
Multiplying both side of the Eq.\ref{eq:cons_Tmunu} by $x_kx_l$ and integrating over all $\textbf{x}$ we get
\begin{eqnarray}
T_{kl}(\omega^{\prime})&=& -\frac{\omega^{\prime}{}^2}{2}\int T_{00}(\textbf{x},\omega^{\prime})x_kx_l d^3x \label{eq:step1}\\
&=& -\frac{\mu\omega^{\prime}{}^2}{2T}\int_0^T\int \delta^3(\textbf{x}^\prime-\textbf{x}(t))e^{i\omega^{\prime}t}x_kx_l d^3x dt \label{eq:step2}\\
&=& -\frac{\mu\omega^{\prime}{}^2}{2T} \int_0^T x^{\prime}_k(t)x^{\prime}_l(t)e^{i\omega^{\prime}t}dt\label{eq:ak2},
\end{eqnarray}
where in the Eq.\ref{eq:step2} we have used the Eq.\ref{eq:7}. Doing integration by parts of Eq.\ref{eq:ak2} and using the Bessel function identities\footnote{$J_{n-1}(z)-J_{n+1}(z)=2J^\prime_n(z), \hspace{0.5cm}J_{n-1}(z)+J_{n+1}(z)=\frac{2n}{z}J_n(z)$}, we can write the different components of stress tensor in the $x-y$ plane. The $xx$-component of stress tensor in the Fourier space is
\bea
T_{xx}(\omega^{\prime})&=&-\frac{\mu\omega^{\prime}{}^2}{2T}\int_0^Tx^2(t)e^{i\omega^{\prime}t}dt \nonumber\\
&=& -\frac{\mu\omega^{\prime}{}^2}{4\pi}\int_0^{2\pi} a^2(\cos \xi -e)^2e^{in\beta}d\beta,
\label{eq:ak2_xx}
\eea
where we have used Eq.\ref{eq:8} and $\omega^{\prime}=n\Omega$, $\beta=\Omega t$, and $T=2\pi/\Omega$. Doing integration by parts of Eq.\ref{eq:ak2_xx} we get
\begin{eqnarray}
T_{xx}(\omega^{\prime})&=& \frac{\mu\omega^{\prime}{}^2}{4\pi i n}\int^{2\pi}_0 e^{i n\beta} \frac{d}{d\beta} (\cos \xi -e)^2\, d\beta \nonumber\\
&=& \frac{\mu\omega^{\prime}{}^2}{2\pi i n}\int^{2\pi}_0\sin \xi (\cos \xi -e) e^{i n\beta} d\xi \nonumber\\
&=& -\frac{\mu\omega^{\prime}{}^2}{8\pi n}\int_0^{2\pi} \left[(e^{2 i \xi}-e^{-2 i \xi})-2e(e^{ i \xi}-e^{-i \xi})\right]e^{i n\beta} d\xi \nonumber\\
&=& -\frac{\mu\omega^{\prime}{}^2}{4 n}\left[J_{n-2}(ne)-2eJ_{n-1}(ne)+2eJ_{n+1}(ne)-J_{n+2}(ne)\right],
\label{eq:Txx}
\end{eqnarray}
where in the last step we used the definition of the Bessel function and $\beta=\Omega t= \xi - e\sin \xi$.

Similarly, 
\bea
T_{yy}(\omega^{\prime})&=&-\frac{\mu\omega^{\prime}{}^2}{2T}\int_0^Ty^2(t)e^{i\omega^{\prime}t}dt \nonumber\\
&=& -\frac{\mu\omega^{\prime}{}^2 (1-e^2)}{4\pi}\int_0^{2\pi} a^2\sin^2\xi \, e^{in\beta}d\beta.
\label{eq:ak2_yy}
\eea

Adding Eq.\ref{eq:ak2_xx} and Eq.\ref{eq:ak2_yy} we get
\begin{eqnarray}
T_{yy}(\omega^{\prime})+ T_{xx}(\omega^{\prime})&=& -\frac{\mu\omega^{\prime}{}^2 a^2}{4\pi}\int^{2\pi}_0 (1-e\cos \xi)^2 e^{in\beta} d\beta \nonumber \\
&=& \frac{\mu \omega^{\prime}{}^{2} a^2 e}{2\pi i n} \int^{2\pi}_0  \sin \xi \, e^{i n\beta}d\beta ,\, \nonumber\\
&=& \frac{\mu \omega^{\prime}{}^{2} a^2 }{2\pi n^2} \int_0^{2\pi} e \cos \xi \, e^{in \beta}d\xi , \, \nonumber\\
&=& \frac{\mu \omega^{\prime}{}^{2} a^2 }{2\pi n^2} \int_0^{2\pi} e^{in \beta}d\xi 
= \frac{\mu \omega^{\prime}{}^{2} a^2 }{n^2}J^2_n(ne). 
\end{eqnarray}

Therefore
\begin{eqnarray}
T_{yy}(\omega^\prime)&=&-T_{xx}(\omega^{\prime})+\frac{\mu \omega^{\prime}{}^{2} a^2 }{n^2}J^2_n(ne)\nonumber\\
&=& \frac{\mu\omega^{\prime2}a^2}{4n}\Big[J_{n-2}(ne)-2eJ_{n-1}(ne)+\frac{4}{n}J_n(ne)+2eJ_{n+1}(ne)-J_{n+2}(ne)\Big].
\end{eqnarray}

The $xy$-component of the Stress Tensor in the Fourier space is
\begin{eqnarray}
T_{xy}(\omega^{\prime})&=&-\frac{\mu\omega^{\prime}{}^2}{2T}\int_0^Tx(t)y(t)e^{i\omega^{\prime}t}dt \nonumber\\
&=& -\frac{\mu\omega^{\prime}{}^2 \sqrt{1-e^2}}{4\pi}\int_0^{2\pi} a^2(\cos \xi -e)\sin \xi \, e^{in\beta}d\beta  \nonumber\\
&=& \frac{\mu\omega^{\prime}{}^2 a^2 \sqrt{1-e^2}}{4\pi i n} \int^{2\pi}_0 \left(\cos(2\xi) -e\cos \xi\right)e^{in\beta}d\xi, \, \nonumber\\
&=& -i\frac{\mu\omega^{\prime}{}^2 a^2 \sqrt{1-e^2}}{4\pi n}\int^{2\pi}_0 \left(\cos(2\xi) -1\right)e^{in\beta}d\xi \nonumber\\
&=& -i\frac{\mu\omega^{\prime}{}^2 a^2 \sqrt{1-e^2}}{4n}\left[J_{n+2}(ne)-2J_n(ne)+J_{n-2}(ne)\right]
\label{eq:ak2_xy}
\end{eqnarray}

For convenience we summarize the final expressions of $T_{ij}(\omega^{\prime})$ as 
\begin{equation}
T_{xx}(\omega^\prime)=-\frac{\mu\omega^{\prime2} a^2}{4n}\Big[J_{n-2}(ne)-2eJ_{n-1}(ne)+2eJ_{n+1}(ne)-J_{n+2}(ne)\Big],\nonumber
\end{equation}
\begin{equation}
T_{yy}(\omega^\prime)=\frac{\mu\omega^{\prime2}a^2}{4n}\Big[J_{n-2}(ne)-2eJ_{n-1}(ne)+\frac{4}{n}J_n(ne)+2eJ_{n+1}(ne)-J_{n+2}(ne)\Big],\nonumber
\end{equation}
\begin{equation}
T_{xy}(\omega^\prime)=\frac{-i\mu\omega^{\prime2}a^2}{4n}(1-e^2)^{\frac{1}{2}}\Big[J_{n-2}(ne)-2J_n(ne)+J_{n+2}(ne)\Big].
\label{eq:13}
\end{equation}
Using Eq.\ref{eq:13}, we get two useful results 
\begin{eqnarray}
T_{ij}(\omega^{\prime})T^{ij*}(\omega^{\prime})&=&\frac{\mu^2\omega^{\prime}{}^4a^4}{8n^2}\Big\{[J_{n-2}(ne)-2eJ_{n-1}(ne)+2eJ_{n+1}(ne)+\frac{2}{n}J_n(ne)-J_{n+2}(ne)]^2 \nonumber\\
&& +(1-e^2)[J_{n-2}(ne)-2J_n(ne)+J_{n+2}(ne)]^2+\frac{4}{n^2}J^2_{n}(ne)\Big\}\nonumber\\
&=& 4\mu^2\omega'^4a^4\left(f(n,e)+\frac{J^2_n(ne)}{12n^4}\right),
\end{eqnarray}
where
\begin{equation}
\begin{split}
f(n,e)=\frac{1}{32n^2}\Big\{[J_{n-2}(ne)-2eJ_{n-1}(ne)+2eJ_{n+1}(ne)+\frac{2}{n}J_n(ne)-J_{n+2}(ne)]^2+\\
(1-e^2)[J_{n-2}(ne)-2J_n(ne)+J_{n+2}(ne)]^2+\frac{4}{3n^2}J^2_{n}(ne)\Big\}
\end{split}
\end{equation}
and
\begin{equation}
\vert T^i{}_i\vert^2= \frac{\mu^2\omega^{\prime}{}^4a^4}{n^4}J^2_{n}(ne).
\end{equation}

 Thus the energy loss due to massless graviton radiation becomes
\begin{eqnarray}
\frac{dE}{dt}&=&\frac{\kappa^2}{8(2\pi)^2}\int\frac{8\pi}{5}\Big[T_{ij}(\omega^\prime)T^*_{ji}(\omega^\prime)-\frac{1}{3}|T^{i}{}_{i}(\omega^\prime)|^2\Big]
\delta(\omega-\omega^\prime)\omega^2 d\omega,\nonumber\\
&=& \frac{32G}{5}\sum^\infty_{n=1}(n\Omega)^2\mu^2a^4(n\Omega)^4f(n,e)\nonumber\\
&=&\frac{32G}{5}\Omega^6\Big(\frac{m_1m_2}{m_1+m_2}\Big)^2a^4(1-e^2)^{-7/2}\Big(1+\frac{73}{24}e^2+\frac{37}{96}e^4\Big).
\label{eq:app10}
\end{eqnarray}
This expression is called Einstein's quadrupole gravitational radiation which matches with the Peters-Mathews result \cite{Peters:1963ux}. From the energy loss formula we can calculate the change in time period ($P_b= 2\pi/\Omega$). From Kepler's law $\Omega^2 a^3= G (M_1+M_2)$ we have $\dot a/a = (2/3)( \dot P_b/P_b) $ . The gravitational energy is $E=-G M_1 M_2/2a$ which implies $\dot a/a= - (\dot E/E)$. Using these two relations we get $\dot P_b/P_b=-(3/2) (\dot E/E)$.

\bibliographystyle{utphys}
\bibliography{bibl}

\providecommand{\href}[2]{#2}\begingroup\raggedright\begin{thebibliography}{10}

\bibitem{Weinberg:1972kfs}
S.~Weinberg, {\em {Gravitation and Cosmology}: {Principles and Applications of
  the General Theory of Relativity}}.
\newblock John Wiley and Sons, New York, 1972.

\bibitem{Fierz:1939ix}
M.~Fierz and W.~Pauli, ``{On relativistic wave equations for particles of
  arbitrary spin in an electromagnetic field},''
  \href{http://dx.doi.org/10.1098/rspa.1939.0140}{{\em Proc. Roy. Soc. Lond. A}
  {\bfseries 173} (1939) 211--232}.

\bibitem{vanDam:1970vg}
H.~van Dam and M.~J.~G. Veltman, ``{Massive and massless Yang-Mills and
  gravitational fields},''
  \href{http://dx.doi.org/10.1016/0550-3213(70)90416-5}{{\em Nucl. Phys. B}
  {\bfseries 22} (1970) 397--411}.

\bibitem{Zakharov:1970cc}
V.~I. Zakharov, ``{Linearized gravitation theory and the graviton mass},'' {\em
  JETP Lett.} {\bfseries 12} (1970) 312.

\bibitem{Hinterbichler:2011tt}
K.~Hinterbichler, ``{Theoretical Aspects of Massive Gravity},''
  \href{http://dx.doi.org/10.1103/RevModPhys.84.671}{{\em Rev. Mod. Phys.}
  {\bfseries 84} (2012) 671--710},
  \href{http://arxiv.org/abs/1105.3735}{{\ttfamily arXiv:1105.3735 [hep-th]}}.

\bibitem{deRham:2014zqa}
C.~de~Rham, ``{Massive Gravity},''
  \href{http://dx.doi.org/10.12942/lrr-2014-7}{{\em Living Rev. Rel.}
  {\bfseries 17} (2014) 7}, \href{http://arxiv.org/abs/1401.4173}{{\ttfamily
  arXiv:1401.4173 [hep-th]}}.

\bibitem{Mitsou:2015yfa}
E.~Mitsou, \href{http://dx.doi.org/10.1007/978-3-319-31729-8}{{\em {Aspects of
  Infrared Non-local Modifications of General Relativity}}}.
\newblock PhD thesis, Geneva U., 2015.
\newblock \href{http://arxiv.org/abs/1504.04050}{{\ttfamily arXiv:1504.04050
  [gr-qc]}}.

\bibitem{Joyce:2014kja}
A.~Joyce, B.~Jain, J.~Khoury, and M.~Trodden, ``{Beyond the Cosmological
  Standard Model},''
  \href{http://dx.doi.org/10.1016/j.physrep.2014.12.002}{{\em Phys. Rept.}
  {\bfseries 568} (2015) 1--98},
  \href{http://arxiv.org/abs/1407.0059}{{\ttfamily arXiv:1407.0059
  [astro-ph.CO]}}.

\bibitem{deRham:2016nuf}
C.~de~Rham, J.~T. Deskins, A.~J. Tolley, and S.-Y. Zhou, ``{Graviton Mass
  Bounds},'' \href{http://dx.doi.org/10.1103/RevModPhys.89.025004}{{\em Rev.
  Mod. Phys.} {\bfseries 89} no.~2, (2017) 025004},
  \href{http://arxiv.org/abs/1606.08462}{{\ttfamily arXiv:1606.08462
  [astro-ph.CO]}}.

\bibitem{Feynman:1996kb}
R.~P. Feynman, {\em {Feynman lectures on gravitation}}.
\newblock 1996.

\bibitem{Weinberg:1964ew}
S.~Weinberg, ``{Photons and Gravitons in $S$-Matrix Theory: Derivation of
  Charge Conservation and Equality of Gravitational and Inertial Mass},''
  \href{http://dx.doi.org/10.1103/PhysRev.135.B1049}{{\em Phys. Rev.}
  {\bfseries 135} (1964) B1049--B1056}.

\bibitem{Veltman:1975vx}
M.~J.~G. Veltman, ``{Quantum Theory of Gravitation},'' {\em Conf. Proc. C}
  {\bfseries 7507281} (1975) 265--327.

\bibitem{Donoghue:2017pgk}
J.~F. Donoghue, M.~M. Ivanov, and A.~Shkerin, ``{EPFL Lectures on General
  Relativity as a Quantum Field Theory},''
  \href{http://arxiv.org/abs/1702.00319}{{\ttfamily arXiv:1702.00319
  [hep-th]}}.

\bibitem{Kuntz:2019zef}
A.~Kuntz, F.~Piazza, and F.~Vernizzi, ``{Effective field theory for
  gravitational radiation in scalar-tensor gravity},''
  \href{http://dx.doi.org/10.1088/1475-7516/2019/05/052}{{\em JCAP} {\bfseries
  05} (2019) 052}, \href{http://arxiv.org/abs/1902.04941}{{\ttfamily
  arXiv:1902.04941 [gr-qc]}}.

\bibitem{Mohanty:1994yi}
S.~Mohanty and P.~Kumar~Panda, ``{Particle physics bounds from the Hulse-Taylor
  binary},'' \href{http://dx.doi.org/10.1103/PhysRevD.53.5723}{{\em Phys. Rev.
  D} {\bfseries 53} (1996) 5723--5726},
  \href{http://arxiv.org/abs/hep-ph/9403205}{{\ttfamily arXiv:hep-ph/9403205}}.

\bibitem{Mohanty:2020pfa}
S.~Mohanty, \href{http://dx.doi.org/10.1007/978-3-030-56201-4}{{\em
  {Astroparticle Physics and Cosmology}: {Perspectives in the Multimessenger
  Era}}}, vol.~975.
\newblock 2020.

\bibitem{Peters:1963ux}
P.~C. Peters and J.~Mathews, ``{Gravitational radiation from point masses in a
  Keplerian orbit},'' \href{http://dx.doi.org/10.1103/PhysRev.131.435}{{\em
  Phys. Rev.} {\bfseries 131} (1963) 435--439}.

\bibitem{Hulse:1974eb}
R.~A. Hulse and J.~H. Taylor, ``{Discovery of a pulsar in a binary system},''
  \href{http://dx.doi.org/10.1086/181708}{{\em Astrophys. J. Lett.} {\bfseries
  195} (1975) L51--L53}.

\bibitem{Taylor:1982zz}
J.~H. Taylor and J.~M. Weisberg, ``{A new test of general relativity:
  Gravitational radiation and the binary pulsar PS R 1913+16},''
  \href{http://dx.doi.org/10.1086/159690}{{\em Astrophys. J.} {\bfseries 253}
  (1982) 908--920}.

\bibitem{Weisberg:1984zz}
J.~M. Weisberg and J.~H. Taylor, ``{Observations of Post-Newtonian Timing
  Effects in the Binary Pulsar PSR 1913+16},''
  \href{http://dx.doi.org/10.1103/PhysRevLett.52.1348}{{\em Phys. Rev. Lett.}
  {\bfseries 52} (1984) 1348--1350}.

\bibitem{Weisberg:2016jye}
J.~M. Weisberg and Y.~Huang, ``{Relativistic Measurements from Timing the
  Binary Pulsar PSR B1913+16},''
  \href{http://dx.doi.org/10.3847/0004-637X/829/1/55}{{\em Astrophys. J.}
  {\bfseries 829} no.~1, (2016) 55},
  \href{http://arxiv.org/abs/1606.02744}{{\ttfamily arXiv:1606.02744
  [astro-ph.HE]}}.

\bibitem{Kramer:2006nb}
M.~Kramer {\em et~al.}, ``{Tests of general relativity from timing the double
  pulsar},'' \href{http://dx.doi.org/10.1126/science.1132305}{{\em Science}
  {\bfseries 314} (2006) 97--102},
  \href{http://arxiv.org/abs/astro-ph/0609417}{{\ttfamily
  arXiv:astro-ph/0609417}}.

\bibitem{Antoniadis:2013pzd}
J.~Antoniadis {\em et~al.}, ``{A Massive Pulsar in a Compact Relativistic
  Binary},'' \href{http://dx.doi.org/10.1126/science.1233232}{{\em Science}
  {\bfseries 340} (2013) 6131},
  \href{http://arxiv.org/abs/1304.6875}{{\ttfamily arXiv:1304.6875
  [astro-ph.HE]}}.

\bibitem{Freire:2012mg}
P.~C.~C. Freire, N.~Wex, G.~Esposito-Farese, J.~P.~W. Verbiest, M.~Bailes,
  B.~A. Jacoby, M.~Kramer, I.~H. Stairs, J.~Antoniadis, and G.~H. Janssen,
  ``{The relativistic pulsar-white dwarf binary PSR J1738+0333 II. The most
  stringent test of scalar-tensor gravity},''
  \href{http://dx.doi.org/10.1111/j.1365-2966.2012.21253.x}{{\em Mon. Not. Roy.
  Astron. Soc.} {\bfseries 423} (2012) 3328},
  \href{http://arxiv.org/abs/1205.1450}{{\ttfamily arXiv:1205.1450
  [astro-ph.GA]}}.

\bibitem{Poddar:2019zoe}
T.~Kumar~Poddar, S.~Mohanty, and S.~Jana, ``{Constraints on ultralight axions
  from compact binary systems},''
  \href{http://dx.doi.org/10.1103/PhysRevD.101.083007}{{\em Phys. Rev. D}
  {\bfseries 101} no.~8, (2020) 083007},
  \href{http://arxiv.org/abs/1906.00666}{{\ttfamily arXiv:1906.00666
  [hep-ph]}}.

\bibitem{Poddar:2019wvu}
T.~Kumar~Poddar, S.~Mohanty, and S.~Jana, ``{Vector gauge boson radiation from
  compact binary systems in a gauged $L_\mu-L_\tau$ scenario},''
  \href{http://dx.doi.org/10.1103/PhysRevD.100.123023}{{\em Phys. Rev. D}
  {\bfseries 100} no.~12, (2019) 123023},
  \href{http://arxiv.org/abs/1908.09732}{{\ttfamily arXiv:1908.09732
  [hep-ph]}}.

\bibitem{Hu:2000ke}
W.~Hu, R.~Barkana, and A.~Gruzinov, ``{Cold and fuzzy dark matter},''
  \href{http://dx.doi.org/10.1103/PhysRevLett.85.1158}{{\em Phys. Rev. Lett.}
  {\bfseries 85} (2000) 1158--1161},
  \href{http://arxiv.org/abs/astro-ph/0003365}{{\ttfamily
  arXiv:astro-ph/0003365}}.

\bibitem{Hui:2016ltb}
L.~Hui, J.~P. Ostriker, S.~Tremaine, and E.~Witten, ``{Ultralight scalars as
  cosmological dark matter},''
  \href{http://dx.doi.org/10.1103/PhysRevD.95.043541}{{\em Phys. Rev. D}
  {\bfseries 95} no.~4, (2017) 043541},
  \href{http://arxiv.org/abs/1610.08297}{{\ttfamily arXiv:1610.08297
  [astro-ph.CO]}}.

\bibitem{Visser:1997hd}
M.~Visser, ``{Mass for the graviton},''
  \href{http://dx.doi.org/10.1023/A:1026611026766}{{\em Gen. Rel. Grav.}
  {\bfseries 30} (1998) 1717--1728},
  \href{http://arxiv.org/abs/gr-qc/9705051}{{\ttfamily arXiv:gr-qc/9705051}}.

\bibitem{Finn:2001qi}
L.~S. Finn and P.~J. Sutton, ``{Bounding the mass of the graviton using binary
  pulsar observations},''
  \href{http://dx.doi.org/10.1103/PhysRevD.65.044022}{{\em Phys. Rev. D}
  {\bfseries 65} (2002) 044022},
  \href{http://arxiv.org/abs/gr-qc/0109049}{{\ttfamily arXiv:gr-qc/0109049}}.

\bibitem{Gambuti:2020onb}
G.~Gambuti and N.~Maggiore, ``{A note on harmonic gauge(s) in massive
  gravity},'' \href{http://dx.doi.org/10.1016/j.physletb.2020.135530}{{\em
  Phys. Lett. B} {\bfseries 807} (2020) 135530},
  \href{http://arxiv.org/abs/2006.04360}{{\ttfamily arXiv:2006.04360 [gr-qc]}}.

\bibitem{Gambuti:2021meo}
G.~Gambuti and N.~Maggiore, ``{Fierz\textendash{}Pauli theory reloaded: from a
  theory of a symmetric tensor field to linearized massive gravity},''
  \href{http://dx.doi.org/10.1140/epjc/s10052-021-08962-8}{{\em Eur. Phys. J.
  C} {\bfseries 81} no.~2, (2021) 171},
  \href{http://arxiv.org/abs/2102.10813}{{\ttfamily arXiv:2102.10813 [gr-qc]}}.

\bibitem{Dvali:2000hr}
G.~R. Dvali, G.~Gabadadze, and M.~Porrati, ``{4-D gravity on a brane in 5-D
  Minkowski space},''
  \href{http://dx.doi.org/10.1016/S0370-2693(00)00669-9}{{\em Phys. Lett. B}
  {\bfseries 485} (2000) 208--214},
  \href{http://arxiv.org/abs/hep-th/0005016}{{\ttfamily arXiv:hep-th/0005016}}.

\bibitem{Dvali:2000rv}
G.~R. Dvali, G.~Gabadadze, and M.~Porrati, ``{Metastable gravitons and infinite
  volume extra dimensions},''
  \href{http://dx.doi.org/10.1016/S0370-2693(00)00631-6}{{\em Phys. Lett. B}
  {\bfseries 484} (2000) 112--118},
  \href{http://arxiv.org/abs/hep-th/0002190}{{\ttfamily arXiv:hep-th/0002190}}.

\bibitem{Dvali:2000xg}
G.~R. Dvali and G.~Gabadadze, ``{Gravity on a brane in infinite volume extra
  space},'' \href{http://dx.doi.org/10.1103/PhysRevD.63.065007}{{\em Phys. Rev.
  D} {\bfseries 63} (2001) 065007},
  \href{http://arxiv.org/abs/hep-th/0008054}{{\ttfamily arXiv:hep-th/0008054}}.

\bibitem{VanNieuwenhuizen:1973qf}
P.~Van~Nieuwenhuizen, ``{Radiation of massive gravitation},''
  \href{http://dx.doi.org/10.1103/PhysRevD.7.2300}{{\em Phys. Rev. D}
  {\bfseries 7} (1973) 2300--2308}.

\bibitem{Will:1997bb}
C.~M. Will, ``{Bounding the mass of the graviton using gravitational wave
  observations of inspiralling compact binaries},''
  \href{http://dx.doi.org/10.1103/PhysRevD.57.2061}{{\em Phys. Rev. D}
  {\bfseries 57} (1998) 2061--2068},
  \href{http://arxiv.org/abs/gr-qc/9709011}{{\ttfamily arXiv:gr-qc/9709011}}.

\bibitem{Larson:1999kg}
S.~L. Larson and W.~A. Hiscock, ``{Using binary stars to bound the mass of the
  graviton},'' \href{http://dx.doi.org/10.1103/PhysRevD.61.104008}{{\em Phys.
  Rev. D} {\bfseries 61} (2000) 104008},
  \href{http://arxiv.org/abs/gr-qc/9912102}{{\ttfamily arXiv:gr-qc/9912102}}.

\bibitem{deRham:2012fw}
C.~de~Rham, A.~J. Tolley, and D.~H. Wesley, ``{Vainshtein Mechanism in Binary
  Pulsars},'' \href{http://dx.doi.org/10.1103/PhysRevD.87.044025}{{\em Phys.
  Rev. D} {\bfseries 87} no.~4, (2013) 044025},
  \href{http://arxiv.org/abs/1208.0580}{{\ttfamily arXiv:1208.0580 [gr-qc]}}.

\bibitem{Shao:2020fka}
L.~Shao, N.~Wex, and S.-Y. Zhou, ``{New Graviton Mass Bound from Binary
  Pulsars},'' \href{http://dx.doi.org/10.1103/PhysRevD.102.024069}{{\em Phys.
  Rev. D} {\bfseries 102} no.~2, (2020) 024069},
  \href{http://arxiv.org/abs/2007.04531}{{\ttfamily arXiv:2007.04531 [gr-qc]}}.

\bibitem{Dvali:2002vf}
G.~Dvali, A.~Gruzinov, and M.~Zaldarriaga, ``{The Accelerated universe and the
  moon},'' \href{http://dx.doi.org/10.1103/PhysRevD.68.024012}{{\em Phys. Rev.
  D} {\bfseries 68} (2003) 024012},
  \href{http://arxiv.org/abs/hep-ph/0212069}{{\ttfamily arXiv:hep-ph/0212069}}.

\bibitem{Talmadge:1988qz}
C.~Talmadge, J.~P. Berthias, R.~W. Hellings, and E.~M. Standish, ``{Model
  Independent Constraints on Possible Modifications of Newtonian Gravity},''
  \href{http://dx.doi.org/10.1103/PhysRevLett.61.1159}{{\em Phys. Rev. Lett.}
  {\bfseries 61} (1988) 1159--1162}.

\bibitem{Fomalont:2009zg}
E.~Fomalont, S.~Kopeikin, G.~Lanyi, and J.~Benson, ``{Progress in Measurements
  of the Gravitational Bending of Radio Waves Using the VLBA},''
  \href{http://dx.doi.org/10.1088/0004-637X/699/2/1395}{{\em Astrophys. J.}
  {\bfseries 699} (2009) 1395--1402},
  \href{http://arxiv.org/abs/0904.3992}{{\ttfamily arXiv:0904.3992
  [astro-ph.CO]}}.

\bibitem{Vainshtein:1972sx}
A.~I. Vainshtein, ``{To the problem of nonvanishing gravitation mass},''
  \href{http://dx.doi.org/10.1016/0370-2693(72)90147-5}{{\em Phys. Lett. B}
  {\bfseries 39} (1972) 393--394}.

\bibitem{Babichev:2013usa}
E.~Babichev and C.~Deffayet, ``{An introduction to the Vainshtein mechanism},''
  \href{http://dx.doi.org/10.1088/0264-9381/30/18/184001}{{\em Class. Quant.
  Grav.} {\bfseries 30} (2013) 184001},
  \href{http://arxiv.org/abs/1304.7240}{{\ttfamily arXiv:1304.7240 [gr-qc]}}.

\bibitem{Arkani-Hamed:2002bjr}
N.~Arkani-Hamed, H.~Georgi, and M.~D. Schwartz, ``{Effective field theory for
  massive gravitons and gravity in theory space},''
  \href{http://dx.doi.org/10.1016/S0003-4916(03)00068-X}{{\em Annals Phys.}
  {\bfseries 305} (2003) 96--118},
  \href{http://arxiv.org/abs/hep-th/0210184}{{\ttfamily arXiv:hep-th/0210184}}.

\bibitem{Poddar:2020exe}
T.~Kumar~Poddar, S.~Mohanty, and S.~Jana, ``{Constraints on long range force
  from perihelion precession of planets in a gauged $L_e-L_{\mu,\tau}$
  scenario},'' \href{http://dx.doi.org/10.1140/epjc/s10052-021-09078-9}{{\em
  Eur. Phys. J. C} {\bfseries 81} no.~4, (2021) 286},
  \href{http://arxiv.org/abs/2002.02935}{{\ttfamily arXiv:2002.02935
  [hep-ph]}}.

\bibitem{KumarPoddar:2021ked}
T.~Kumar~Poddar, ``{Constraints on axionic fuzzy dark matter from light bending
  and Shapiro time delay},'' \href{http://arxiv.org/abs/2104.09772}{{\ttfamily
  arXiv:2104.09772 [hep-ph]}}.

\bibitem{TheLIGOScientific:2016src}
{\bfseries LIGO Scientific, Virgo} Collaboration, B.~P. Abbott {\em et~al.},
  ``{Tests of general relativity with GW150914},''
  \href{http://dx.doi.org/10.1103/PhysRevLett.116.221101}{{\em Phys. Rev.
  Lett.} {\bfseries 116} no.~22, (2016) 221101},
  \href{http://arxiv.org/abs/1602.03841}{{\ttfamily arXiv:1602.03841 [gr-qc]}}.
  [Erratum: Phys.Rev.Lett. 121, 129902 (2018)].

\bibitem{Boulware:1973my}
D.~G. Boulware and S.~Deser, ``{Can gravitation have a finite range?},''
  \href{http://dx.doi.org/10.1103/PhysRevD.6.3368}{{\em Phys. Rev. D}
  {\bfseries 6} (1972) 3368--3382}.

\bibitem{Dvali:2006su}
G.~Dvali, ``{Predictive Power of Strong Coupling in Theories with Large
  Distance Modified Gravity},''
  \href{http://dx.doi.org/10.1088/1367-2630/8/12/326}{{\em New J. Phys.}
  {\bfseries 8} (2006) 326},
  \href{http://arxiv.org/abs/hep-th/0610013}{{\ttfamily arXiv:hep-th/0610013}}.

\bibitem{Deffayet:2001pu}
C.~Deffayet, G.~R. Dvali, and G.~Gabadadze, ``{Accelerated universe from
  gravity leaking to extra dimensions},''
  \href{http://dx.doi.org/10.1103/PhysRevD.65.044023}{{\em Phys. Rev. D}
  {\bfseries 65} (2002) 044023},
  \href{http://arxiv.org/abs/astro-ph/0105068}{{\ttfamily
  arXiv:astro-ph/0105068}}.

\bibitem{Rubakov:2004eb}
V.~A. Rubakov, ``{Lorentz-violating graviton masses: Getting around ghosts, low
  strong coupling scale and VDVZ discontinuity},''
  \href{http://arxiv.org/abs/hep-th/0407104}{{\ttfamily arXiv:hep-th/0407104}}.

\bibitem{Dubovsky:2004sg}
S.~L. Dubovsky, ``{Phases of massive gravity},''
  \href{http://dx.doi.org/10.1088/1126-6708/2004/10/076}{{\em JHEP} {\bfseries
  10} (2004) 076}, \href{http://arxiv.org/abs/hep-th/0409124}{{\ttfamily
  arXiv:hep-th/0409124}}.

\bibitem{Rubakov:2008nh}
V.~A. Rubakov and P.~G. Tinyakov, ``{Infrared-modified gravities and massive
  gravitons},'' \href{http://dx.doi.org/10.1070/PU2008v051n08ABEH006600}{{\em
  Phys. Usp.} {\bfseries 51} (2008) 759--792},
  \href{http://arxiv.org/abs/0802.4379}{{\ttfamily arXiv:0802.4379 [hep-th]}}.

\bibitem{Kostelecky:2016kfm}
V.~A. Kosteleck\'y and M.~Mewes, ``{Testing local Lorentz invariance with
  gravitational waves},''
  \href{http://dx.doi.org/10.1016/j.physletb.2016.04.040}{{\em Phys. Lett. B}
  {\bfseries 757} (2016) 510--514},
  \href{http://arxiv.org/abs/1602.04782}{{\ttfamily arXiv:1602.04782 [gr-qc]}}.

\bibitem{Kostelecky:2016uex}
V.~A. Kosteleck\'y and M.~Mewes, ``{Testing local Lorentz invariance with
  short-range gravity},''
  \href{http://dx.doi.org/10.1016/j.physletb.2016.12.062}{{\em Phys. Lett. B}
  {\bfseries 766} (2017) 137--143},
  \href{http://arxiv.org/abs/1611.10313}{{\ttfamily arXiv:1611.10313 [gr-qc]}}.

\bibitem{TheLIGOScientific:2014jea}
{\bfseries LIGO Scientific} Collaboration, J.~Aasi {\em et~al.}, ``{Advanced
  LIGO},'' \href{http://dx.doi.org/10.1088/0264-9381/32/7/074001}{{\em Class.
  Quant. Grav.} {\bfseries 32} (2015) 074001},
  \href{http://arxiv.org/abs/1411.4547}{{\ttfamily arXiv:1411.4547 [gr-qc]}}.

\bibitem{TheVirgo:2014hva}
{\bfseries VIRGO} Collaboration, F.~Acernese {\em et~al.}, ``{Advanced Virgo: a
  second-generation interferometric gravitational wave detector},''
  \href{http://dx.doi.org/10.1088/0264-9381/32/2/024001}{{\em Class. Quant.
  Grav.} {\bfseries 32} no.~2, (2015) 024001},
  \href{http://arxiv.org/abs/1408.3978}{{\ttfamily arXiv:1408.3978 [gr-qc]}}.

\bibitem{Cardoso:2018zhm}
V.~Cardoso, G.~Castro, and A.~Maselli, ``{Gravitational waves in massive
  gravity theories: waveforms, fluxes and constraints from extreme-mass-ratio
  mergers},'' \href{http://dx.doi.org/10.1103/PhysRevLett.121.251103}{{\em
  Phys. Rev. Lett.} {\bfseries 121} no.~25, (2018) 251103},
  \href{http://arxiv.org/abs/1809.00673}{{\ttfamily arXiv:1809.00673 [gr-qc]}}.

\end{thebibliography}\endgroup
\end{document}